\documentclass{emulateapj}
\usepackage{epsfig}

\newcommand{\beq}{\begin{equation}}
\newcommand{\eeq}{\end{equation}}

\def\be{\begin{equation}}
\def\ee{\end{equation}}
\def\bea{\begin{eqnarray}}
\def\eea{\end{eqnarray}}



\def \logTd6 {\hbox{log$( T/6 \kev)$} }

\def\myputfigure#1#2#3#4#5%
{\vskip#5pt\makebox[0pt]{\hskip#2in
\includegraphics[width=#3\textwidth]{#1}}\vskip#4pt\hfill}

\def \arcsec    {^{\prime\prime}}

\def \kms            {~{\rm km~s}^{-1}}

\def \etal      {et al.}

\def \kev       {{\rm\ keV}}
\def \msol      {{\rm\ M}_\odot}

\def \hMpc      {h^{-1}{\rm\ Mpc}}
\def \hkpc      {h^{-1}{\rm\ kpc}}

\input epsf
\bibpunct{(}{)}{;}{a}{}{,}

\begin{document}

\lefthead{BINARY QUASARS}
\righthead{Hennawi \etal}

\title{Binary Quasars in the Sloan Digital Sky Survey: Evidence for Excess Clustering on Small Scales}

\author{Joseph F. Hennawi,\altaffilmark{1,2,3} 
  Michael A. Strauss,\altaffilmark{3}
  Masamune Oguri,\altaffilmark{3,4} 
  Naohisa Inada,\altaffilmark{5}
  Gordon T. Richards,\altaffilmark{3} 
  Bartosz Pindor,\altaffilmark{6,7}
  Donald P. Schneider,\altaffilmark{8} 
  Robert H. Becker,\altaffilmark{9,10}
  Michael D. Gregg,\altaffilmark{9,10}
  Patrick B. Hall,\altaffilmark{11}
  David E. Johnston,\altaffilmark{3}
  Xiaohui Fan,\altaffilmark{12}
  Scott Burles,\altaffilmark{13}
  David J. Schlegel,\altaffilmark{14}
  James E. Gunn,\altaffilmark{3}
  Robert Lupton,\altaffilmark{3}
  Neta A. Bahcall,\altaffilmark{3}
  Robert J. Brunner,\altaffilmark{15}
  Jon Brinkman\altaffilmark{16}
}

\altaffiltext{1}{Hubble Fellow}
\altaffiltext{2}{Department of Astronomy, University of California at
Berkeley, 601 Campbell Hall, Berkeley, CA 94720-3411.}
\altaffiltext{3}{Princeton University Observatory, Peyton Hall,
  Princeton, NJ 08544.}
\altaffiltext{4}{Department of Physics, University of Tokyo, Hongo
7-3-1, Bunkyo-ku, Tokyo 113-0033, Japan.}
\altaffiltext{5}{Institute of Astronomy, University of Tokyo, 2-21-1
Osawa, Mitaka, Tokyo 181-8588, Japan.}
\altaffiltext{6}{Department of Astronomy, University of Toronto, 60 St George Street, Toronto M5S3H8, Canada}
\altaffiltext{7}{CFHT Legacy Survey Postdoctoral Fellow}
\altaffiltext{8}{Department of Astronomy and Astrophysics, Pennsylvania State
University, 525 Davey Laboratory, University Park, PA 16802.}
\altaffiltext{9}{Department of Physics, University of California at
Davis, 1 Shields Avenue, Davis, CA 95616.}
\altaffiltext{10}{Institute of Geophysics and Planetary Physics,
Lawrence Livermore National Laboratory, L-413, 7000 East Avenue,
Livermore, CA 94550.}
\altaffiltext{13}{Department of Physics \& Astronomy,
York University, 4700 Keele St., Toronto, ON M3J 1P3, Canada.}
\altaffiltext{12}{Steward Observatory, University of Arizona, 933 North Cherry Avenue, Tucson, AZ 85721}
\altaffiltext{13}{Physics Department, Massachusetts Institute of Technology, 77 Massachusetts Avenue, Cambridge, MA 02139.} 
\altaffiltext{14}{Lawrence Berkeley National Laboratory, One Cyclotron Road, Mailstop 50R232, Berkeley, CA, 94720, USA.}
\altaffiltext{15}{Department of Astronomy and National Center for Supercomputer Applications, University of Illinois, 1002 West Green Street, Urbana, IL 61801.}
\altaffiltext{16}{Apache Point Observatory, P. O. Box 59, Sunspot, NM88349-0059.}

\begin{abstract}
  We present a sample of 218 new quasar pairs with proper transverse
  separations $R_{\rm prop} < 1~\hMpc$ over the redshift range $0.5 <
  z < 3.0$, discovered from an extensive follow up campaign to find
  companions around the Sloan Digital Sky Survey and 2dF Quasar
  Redshift Survey quasars. This sample includes 26 new binary quasars
  with separations $R_{\rm prop}< 50~\hkpc$ ($\theta < 10\arcsec$),
  more than doubling the number of such systems known. We define a
  statistical sample of binaries selected with homogeneous criteria
  and compute its selection function, taking into account sources of
  incompleteness. The first measurement of the quasar correlation
  function on scales $10~\hkpc < R_{\rm prop}< 400~\hkpc$ is
  presented. For $R_{\rm prop}\lesssim 40~\hkpc$, we detect an order
  of magnitude excess clustering over the expectation from the large
  scale ($R_{\rm prop}\gtrsim 3~\hMpc$) quasar correlation function,
  extrapolated down as a power law to the separations probed by our
  binaries. The excess grows to $\sim 30$ at $R_{\rm prop}\sim
  10~\hkpc$, and provides compelling evidence that the quasar
  autocorrelation function gets progressively steeper on sub-Mpc
  scales.  This small scale excess can likely be attributed to 
  dissipative interaction events which trigger quasar activity in rich
  environments. Recent small scale measurements of galaxy clustering
  and quasar-galaxy clustering are reviewed and discussed in relation
  to our measurement of small scale quasar clustering.
\end{abstract}


\section{Introduction}
\label{sec:intro}

A fundamental problem for cosmologists is to understand how quasars
are embedded in the galaxy formation hierarchy and to relate them to
the gravitational evolution of the structure of the underlying
dark matter. In the current paradigm, every massive galaxy is thought
to have undergone a luminous quasar phase, and quasars at high redshift
are the progenitors of the local dormant supermassive black hole
population found in the centers of nearly all nearby bulge-dominated galaxies
\citep[e.g.][]{SB92,YT02} . This fundamental connection is supported
by the tight correlations between the masses of central black holes
and the velocity dispersions of their old stellar populations
\citep{Mag98,FM00,Geb00,Trem02} and by comparing the number density of black 
holes in the local Universe to the luminosity density produced by 
quasars at high redshift \citep{SB92,YT02}.  

Quasars are likely to reside in massive hosts \citep{Turner91} and it
has been suggested that they occupy the rarest peaks in the initial
Gaussian density fluctuation distribution
\citep{ER88,CK89,NS93,Djor99a,Djor99b,Djor03,Stia05}. It is also thought that
quasar activity is triggered by the frequent mergers which are a
generic consequence of bottom up structure formation models
\citep{Bahcall97,Carl90,HR93,WL02}. Both of these hypothesis imply
that quasars should be highly biased tracers of the dark matter
distribution: rare peaks in the density field are intrinsically
strongly clustered \citep{Kaiser84} and the frequency of mergers is
higher in dense environments \citep{LC93}. Measurements of quasar
clustering can thus teach us about the environments of quasars as well 
as give clues to the dynamical processes which trigger quasar activity.
Furthermore, a comparison of quasar clustering with the quasar
luminosity function can be used to constrain the mean quasar lifetime
\citep{HH01,MW01} as well as the relationship between the mass of
central black holes and the circular velocities of their host dark
halos \citep{WL04}. 

There have been many attempts to measure quasar clustering, beginning
with the pioneering work of \citet{Osmer81}. \citet{Shaver84} first
detected quasar clustering using a clever technique to measure
correlations from inhomogeneous catalogs and discovered that quasars
were clustered similar to galaxies in the local universe, a result
later confirmed by \citet{Kru88}.  Most recently, \citet{Croom01},
\citet{Porci04} (henceforth PMN), and \citet{Croom05} measured the
clustering of quasars in the redshift range $z=0.3-2.2$ from the $\sim
15,000$ quasars in the Two Degree Field Quasar Survey (2QZ)
\citep{Croom04}. They both find good agreement with a power law
correlation function $\xi(r)=(r/r_0)^{-\gamma}$ on scales
$r=1-35~\hMpc$, with correlation length $r_0\sim 5~\hMpc$ (comoving)
and slope $\gamma\sim 1.5$, with only a weak dependence on redshift 
and luminosity \citep{Croom02} .  This agrees with previous measurements
\citep{IS88,AC92,MF93,SB94,CS96} and is similar to the clustering of
nearby galaxies.

At redshift $z> 2.5$, quasars are rarer and thus quasar clustering has
been much more difficult to measure.  However the mere existence of a
few high redshift quasar pairs provides circumstantial evidence that
quasars may have been much more highly clustered in the past.  In the
Palomar Transit Grism Survey of \citet{SSG94}, three quasar pairs with
$z\gtrsim 3$ and comoving separations $5-10~\hMpc$ were found in a
complete sample of 90 quasars covering 61.5 deg$^2$. Analysis of
clustering in this high redshift sample by \citet{Kundic97} and
\citet{Stephens97} detected a statistically significant clustering
signal, dominated by the three pairs, which implied a comoving correlation
length $r_{0}\sim 50 \hMpc$. This is much larger than the correlation
length of present day galaxies or $z\sim 1.5$ quasars. The only
sub-arcminute high redshift quasar pair known is a $33\arcsec$ pair of
quasars at $z=4.25$ discovered serendipitously by
\citet{Schneider00}. Based on the discovery of this one object with
proper transverse separation of $162~\hkpc$, they estimated the
correlation length could be as large as $\sim
30~\hMpc$. \citet{Djor03} discovered a companion at $z=5.02$ separated
by $196\arcsec$ from the high redshift quasar at $z=4.96$ discovered
by \citet{Fan99}, corresponding to a proper transverse separation of
$896~\hkpc$. This is the highest redshift pair of quasars known.

Even in the large quasar sample studied by \citet{Croom01}, PMN, and
\citet{Croom05}, the smallest scale at which the correlation function
can be measured is $\sim 1~\hMpc$. The reason for this is
twofold. First, close quasar pairs with angular separations $\lesssim
60\arcsec$, corresponding to $\sim 1 \hMpc$ at $z\sim 1.5$, are
extremely rare, simply because at small separations, the correlation
function does not increase as fast as the volume decreases.  Second,
because of the finite size of the optical fibers of the 2dF
multi-object spectrograph, only one quasar in a close pair with
separation $<30\arcsec$ can be observed. This limitation, referred to
as a \emph{fiber collision}, is also a problem for the Sloan Digital
Sky Survey \citep[SDSS]{York00}, for which this angular scale is $55\arcsec$
\citep{Blanton03}.


A significant motivation for studying small scale quasar clustering,
is the existence of controversial population of quasar pairs
discovered in the search for gravitationally lensed quasars
\citep{Koch99,MWF99}. These close pairs have similar optical spectra,
small velocity differences, and typically have separations in the
range $2\arcsec \lesssim \Delta\theta \lesssim 10\arcsec$
characteristic of group or cluster scale lenses. However, deep imaging
shows no identifiable lenses in the foreground. Although a handful of
wide separation ($\Delta \theta > 3\arcsec$) gravitational lenses have
been discovered \citep{Walsh79}, especially recently in the SDSS
\citep{Quad03,Quad04,Oguri05}, the expected number of quasars lensed by
groups and clusters is too small to account for all of the
controversial pairs \citep{OK04,Hennawi05a}. The poster child example
is Q~2345+007 \citep{Weedman82}, the famous pair of $z=2.16$ quasars
with $7.1\arcsec$ separation.  Although a plethora of papers can be
found in the literature arguing for
\citep{Weedman82,Turner82,SS91,Bonnet93,Fischer94,Pello96,Small97}, or
against \citep{PB86,Djor91,Schneider93,Koch99,MWF99,Green02} the
lensing hypothesis for this system, the most compelling argument is
based on the recent Chandra observations of \citet{Green02} who failed
to detect diffuse X-ray emission associated with the potential lens.

This population has thus led to much speculation about exotic mass
concentrations which could be responsible for the apparent multiple
imaging. It has been suggested that the lenses in these systems are
`dark' galaxies or galaxy clusters
\citep{Sub87,Duncan91,Hawkins97,Mal97,Peng99,Koop00}, that they could be
lensed by free floating $\sim 10^{14} \msol$ black holes
\citep{Turner91}, or that they might be instances of gravitational
lensing by cosmic strings \citep{Vil84,Pac86,HN84}.

A much more plausible explanation is that these controversial pairs
are binaries rather than lenses
\citep{PB86,Djor91,Schneider93,Koch99,MWF99} and hence just a
manifestation of quasar clustering on small scales.  \citet{Djor91}
first pointed out that this interpretation implies a factor of $\sim
100$ more binary quasars over what is naively expected from
extrapolating the quasar correlation function power law down to
comoving scales $\lesssim 100~\hkpc$, and he proposed that this was
due to the enhancement of quasar activity during merger events. Based
on two close pairs found in the LBQS survey, \citet{Hew98} similarly
claimed an excess of $\sim 100$ over the expectation from quasar
clustering. \citet{Koch99} compared the optical and radio properties
of the controversial quasar pairs, and argued that they were all binary
quasars, and similarly claimed that the excess binaries could be
explained in a merger scenario. 

The study of binary quasars and small scale quasar clustering has been
hindered by the small number of known examples and the heterogeneous
mix of detection methods.  In this paper we conduct a systematic
search for binary quasars in the SDSS and 2QZ quasar samples. We
present a sample of 218 new binary quasars with proper transverse 
separations $R_{\rm prop}<1~\hMpc$, 24 of which have
angular separations $\lesssim 10\arcsec$ corresponding to transverse
proper separations $R_{\rm prop}\lesssim 50\hkpc$, more than doubling
the number of such systems known. A sub-sample of pairs selected with
well defined criteria is constructed and we quantify its selection
function.  Based on this sample, we present the first measurement of the
correlation function of quasars on the small scales $10~\hkpc \lesssim
R_{\rm prop} \lesssim 1~\hMpc$. We detect excess small scale
clustering compared to the expectation from an extrapolation of the
larger scale two point correlation function power law slope.

The outline of this paper is as follows. In \S \ref{sec:selection}, we
discuss color-selection criteria used to find binary quasars and
describe the follow-up observations required to confirm quasar pair
candidates in \S~\ref{sec:APO}.  Our binary quasar sample is presented in \S
\ref{sec:sample}.  We show that the number of binary quasars
discovered thus far in the SDSS imply an excess of small scale quasar
clustering in \S \ref{sec:clustering} and we compare this result to
small scale galaxy and quasar-galaxy clustering in \S
\ref{sec:qso_gal}.  We summarize and conclude in \S
\ref{sec:conc}. In the Appendix, we present tables summarizing the
results of all of our follow-up observations, as well as a catalog of
projected quasar pairs from the SDSS.

Throughout this paper we use the best fit WMAP (only) cosmological
model of \citet{Spergel03}, with $\Omega_m = 0.270$, $\Omega_\Lambda
=0.73$, $h=0.72$.  Because both proper and comoving distances are
used, we will always indicate the former as $R_{\rm prop}$. It is
helpful to remember that in the chosen cosmology, for a typical quasar
redshift of $z=1.5$, an angular separation of $\Delta\theta=1\arcsec$
corresponds to a proper (comoving) transverse separation of $R_{\rm
  prop}= 6~\hkpc$ ($R= 15~\hkpc$), and a velocity difference of
$1000~\kms$ at this redshift corresponds to a proper radial redshift
space distance of $s_{\rm prop}=4.4~\hMpc$ (comoving $s= 11~\hMpc$). 

\section{Quasar Samples}
\label{sec:samples}

In this section we present a variety of techniques used to select
quasar pair candidates. First, we describe the quasar catalogs which
served as the parent samples for our quasar pair search. Then we
introduce a statistic which quantifies the color similarity of two
quasars. Finally, we discuss each selection method in detail and
describe our follow-up observations.

\subsection{The SDSS Spectroscopic Quasar Sample}

The Sloan Digital Sky Survey uses a dedicated 2.5m telescope and a
large format CCD camera \citep{Gunn98} at the Apache Point
Observatory in New Mexico to obtain images in five broad bands \citep[$u$,
$g$, $r$, $i$ and $z$, centered at 3551, 4686, 6166, 7480 and 8932
\AA, respectively;][]{Fuku96,Stoughton02} of high Galactic latitude sky in the
Northern Galactic Cap.  The imaging data are processed by the
astrometric pipeline \citep{Astrom} and photometric pipeline
\citep{Photo}, and are photometrically calibrated to a standard star
network \citep{Smith02,Hogg01}. Additional details on the SDSS data products
can be found in \citet{DR1,DR2,DR3}.

Based on this imaging data, spectroscopic targets chosen by various
selection algorithms (i.e. quasars, galaxies, stars, serendipity) are
observed with two double spectrographs producing spectra covering
\hbox{3800--9200 \AA} with a spectral resolution ranging from 1800 to
2100.  Details of the spectroscopic observations can be found in
\citet{York00}, \citet{Castander01}, and \citet{Stoughton02}.  A
discussion of quasar target selection can be found in
\citet{qsoselect}.  The Third Data Release Quasar Catalog contains
46,420 quasars \citep{Schneider05}.  Here, we use a larger sample of
quasars, as we include non-public data: our parent
sample includes 67,385 quasars with
$z>0.3$, of which 52,279 quasars lie in the redshift range $0.7 < z <
3.0$. Note 
also that we have used the Princeton/MIT spectroscopic
reductions\footnote{Available at http://spectro.princeton.edu} which
differ slightly from the official SDSS data release.

Most quasar candidates are selected based on their location in
multidimensional SDSS color-space. All magnitudes are reddening
corrected following the prescription in \citet{SFD98}. Objects with
colors that place them outside of the stellar locus which do not
inhabit specific ``exclusion" regions (e.g., places dominated by white
dwarfs, A stars, and M star-white dwarf pairs) are identified as
primary quasar candidates.  An $i$ magnitude limit of~19.1 is imposed
for candidates whose colors indicate a probable redshift of less
than~$\approx$~3; high-redshift candidates are accepted if \hbox{$i <
  20.2$.}  Over 90\% of SDSS-selected quasars follow a remarkably
tight color-redshift relation in the SDSS color-system
\citep{Richards01}. In addition to the multicolor selection,
unresolved objects brighter \hbox{than $i = 19.1$} that lie
within~1.5$''$ of a FIRST radio source \citep{BWH95} are also
identified as primary quasar candidates.

Supplementing the primary quasar sample described above are quasars
targeted by other SDSS target selection packages: Galaxy \citep[the
  SDSS main and extended galaxy samples][]{Eis01,Strauss02}, X-ray
\citep[objects near the position of a ROSAT All-Sky Survey
  source,][]{Anderson03}, Star (point source with unusual color), or
Serendipity (unusual color or FIRST matches).  No attempt at
completeness is made for the last three categories; objects selected
by these algorithms are observed if a given spectroscopic plate has
fibers remaining after all of the high-priority classes (galaxies,
quasars, and sky and spectrophotometric calibrations) in the field
have been assigned fibers \citep[see][]{Blanton03}.  Most of the
quasars that fall below the magnitude limits of the quasar survey were
selected by the serendipity algorithm.

As we described in the introduction, the SDSS spectroscopic survey
selects \emph{against} close pairs of quasars because of fiber
collisions.  The finite size of optical fibers implies only one quasar
in a pair with separation $<55\arcsec$ can receive a fiber.  Follow-up
spectroscopy is thus required to discover quasar pairs.  An exception
to this rule exists for a fraction ($\sim 30\%$) of the area of the
spectroscopic survey covered by overlapping plates.  For these plates
the same area of sky was observed spectroscopically on more than one
occasion so that there is no fiber collision limitation. 

\subsection{The SDSS Faint Photometric Quasar Sample}

\citet{Richards04} have demonstrated that faint ($i\lesssim
21$) photometric samples of quasars can be constructed from the SDSS
photometry, by separating quasars from stars using knowledge of their
relative densities in color space. Each member of this catalog is
assigned a probability of being a quasar, a photometric redshift, and
a probability that the photometric redshift is correct (see
\citealt{Richards04} for details).  We searched for (and found) quasar
pairs in a 
photometric sample of 273,287 quasar \emph{candidates}.  Note that the
faint photometric quasar used here is based on the larger SDSS Data
Release 3 area (DR3; \citealt{DR3}), whereas that published in
\citep{Richards04} 
covers the smaller SDSS DR1 area \citep{DR1}.

\subsection{The SDSS+2QZ Quasar Sample}

The 2dF Quasar Redshift Survey (2QZ) is a homogeneous spectroscopic
catalog of 44,576 stellar objects with 18.25 $\leq b_J\leq$ 20.85
\citep{Croom04}. Of these, 23,338 are quasars spanning the redshift
range $0.3\lesssim z \lesssim 2.9$. Selection of quasar candidates is
based on broad band colors $(u b_J r)$ from automated plate
measurements of the United Kingdom Schmidt Telescope 
photographic plates. Spectroscopic observations were carried out with
the 2dF instrument, which is a multi-object spectrograph at the
Anglo-Australian Telescope. The 2QZ covers a total area of 721.6
deg$2$ arranged in two $75^{\circ}\times 5^{\circ}$ strips across the
South Galactic Cap (SGP strip), centered on $\delta = -30^{\circ}$,
and North Galactic Cap (NGP strip, or equatorial strip), centered at
$\delta=0^{\circ}$. The NGP overlaps the SDSS footprint,
corresponding to roughly half of the 2QZ area. 

By combining the SDSS quasar catalog with 2QZ quasars in the NGP
which have matching SDSS photometry, we arrive at a combined sample of 
75,579 quasars with $z>0.3$, of which 67,385 are from the SDSS and 8194 
from the 2QZ. For the clustering analysis in \S~\ref{sec:clustering} we 
will restrict attention to the redshift range $0.7 < z < 3.0$, for which 
we define a  combined SDSS/2QZ sample of 59,608 quasars, of which 
52,279 are from the SDSS and 6879 are from the 2QZ. 


\section{Quasar Pair Selection}
\label{sec:selection}

Several different techniques are used to find binary quasars.  For
small separation pairs, $\Delta\theta\leq3\arcsec$, characteristic of
the majority of gravitational lenses, binary quasars were discovered
in the SDSS search for gravitationally lensed quasars (e.g.,
\citealt{Oguri05}).  For wider 
separations, both components are resolved and we can exploit the
precise digital photometry of the SDSS to color select quasar pair
candidates. Finally, quasar pairs can be found directly from the
spectroscopy: the SDSS contains a fair number of overlapping
spectroscopic plates for which fiber collision does not limit the pair
separation and quasar pairs can be found by searching the SDSS+2QZ
quasar catalog over regions where the survey areas overlap. 

\subsection{Lens Selection}
\label{sec:lens_selection}

For small separation pairs $\Delta\theta\leq3\arcsec$, the two images
are unresolved or marginally resolved, as the SDSS imaging has a
median seeing of $1.4\arcsec$. Candidates are selected by fitting a
multi-component PSF model to atlas images of each of the SDSS quasars
as described in \citet{Pindor03} and \citet{Inada03}.  We restricted
attention to candidates in the redshift range $0.7<z<3.0$. Quasars
with $z<0.7$ are unlikely to be gravitational lenses, and the PSF
fitting is complicated by the presence of resolved host galaxy
emission. The SDSS is biased against gravitational lenses with $z>3.0$
because the target selection algorithm for high-z quasars does not
target objects classified as extended by the photometric pipeline, and
most candidate lenses and binaries appear extended. See the
discussion in \citet{Pindor03} for more details. The number of quasars
in this redshift range searched with our lens algorithm 
was 39,142, which makes up the parent sample of our lens
search. This is a subset of the total number (52,279) of 
SDSS quasars in this range, as the lens algorithm was 
run on a sample of quasars defined at an earlier date.  We refer to objects
selected by this algorithm as the `lens' sample. 

Follow up observations are required to confirm that the
companion object is indeed a quasar at the same redshift. The limiting
magnitude for the companion objects is $i<21.0$ (here and throughout
we always quote reddening corrected asinh magnitudes
\citep{Lupton99}), as fainter objects are too difficult to observe
from the 3.5m telescope at Apache Point Observatory (APO), where most
of the follow-up observations were conducted, even in the best
conditions (see \S~\ref{sec:APO}).

\subsection{$\chi^2$ Color Selection}

\begin{figure}
  \centerline{
    \epsfig{file=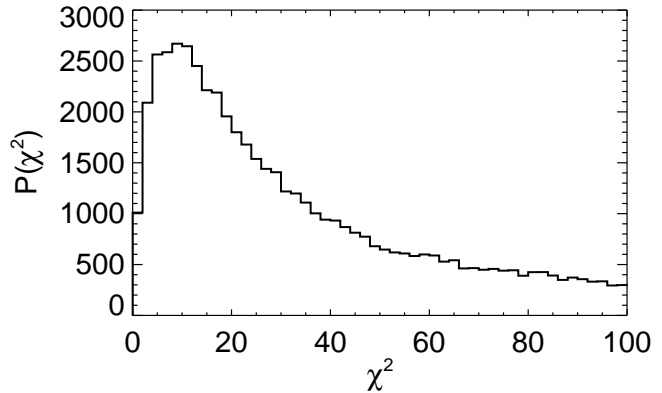,bb= 20 0 504 330,width=0.5\textwidth}}
  \caption{
    Distribution of the $\chi^2$ statistic for the 64621 unique pair
    combinations of 359 quasars in the SDSS sample in the redshift
    interval $2.4<z<2.45$. The median value of this distribution is
    $33.1$, so that a quasar pair survey which aims to achieve 50\%
    completeness in this redshift interval would have to observe all
    quasar pair candidates with $\chi^2<33.1$. 
    \label{fig:chi2_dist}}
\end{figure}

\begin{figure}
  \centerline{
    \epsfig{file=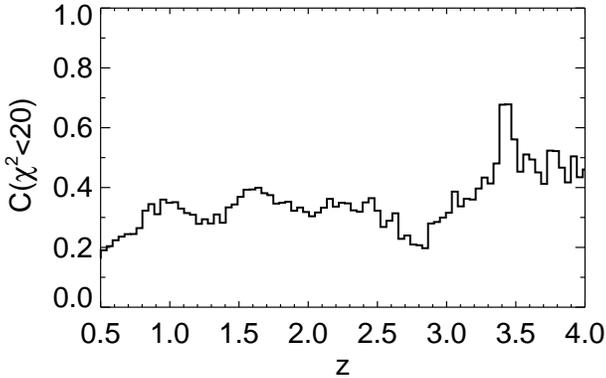,bb=10 0 504 320,width=0.5\textwidth}}
  \caption{ Completeness of $\chi^2$ statistic as a function of
    redshift for a survey which observes all quasar pair candidates
    with $\chi^2<20$.  Dispersion in the color-redshift relation of
    quasars \citep{Richards01} gives rise to a broad distribution of
    $\chi^2$ for pairs of quasars at the same redshift (see
    Figure~\ref{fig:chi2_dist}). Restricting the observations of pair
    candidates to those with $\chi^2<20$ results in $\sim 35\%$
    completeness in the redshift range $0.70<z<3.0$.
    \label{fig:chi2_complete}}
\end{figure}

Although quasars have a wide range of luminosities, the majority have
similar optical/ultraviolet spectral energy
distributions. \citep{Richards01} demonstrated that most quasars
follow a relatively tight color-redshift relation in the SDSS filter
system; a property which has been exploited to calculate photometric
redshifts of quasars \citep{photoz,Buda01,Weinstein04}. It is thus 
possible to efficiently select pairs of quasars at the same redshift
by searching for pairs of objects with similar, quasar-like colors. 

To this end, we define a statistic that quantifies the likelihood that
two astronomical objects have the same color. Recall that a color $u-g$, 
is a statement about flux \emph{ratios}, $f^{u}\slash f^{g}$. Thus if two 
objects have the same color, then their fluxes should be proportional. Given
the fluxes $f^m_1$ of the first, we can ask whether the fluxes of the second
are consistent with the proportionality 
\beq
f^m_2 = A~f^m_1 ,
\label{eqn:model}
\eeq
where $f^m$ is a five dimensional vector of fluxes (one for each SDSS band), 
$m$ designates the filter, and this relationship holds with the same 
proportionality constant $A$ in all bands.

The maximum likelihood value of the parameter A, given the fluxes of both 
objects, can be determined by minimizing the $\chi^2$
\beq
\chi^2(A) = \sum_{\rm ugriz}\frac{(f^m_2- A f^m_1)^2}
    {[\sigma^m_2]^2 + A^2 [\sigma^m_1]^2}, \label{eqn:chi}
\eeq
where we have dropped a term corresponding to the normalization of the 
likelihood because of its slow variation with the parameter $A$. 
The $\sigma^m$ are the photometric measurement errors, but do not
include the intrinsic scatter about the mean color-redshift relation
(see discussion below). We thus arrive at the implicit equation for $A$
\beq
A = \frac{\sum_{\rm ugriz} \frac{f^m_2f^m_1}
{[\sigma^m_2]^2 + A^2 [\sigma^m_1]^2}}{\sum_{\rm ugriz} \frac{[f^m_2]^2}
{[\sigma^m_2]^2 + A^2 [\sigma^m_1]^2}}, 
\eeq
which can be solved in a few iterations.  This value is 
inserted into eqn.~(\ref{eqn:chi}), reducing the
number of degrees of freedom in the $\chi^2$ to four, i.e. the number
of independent colors one could have formed from the five
magnitudes. 

If the fluxes of the two objects are proportional and if the
photometric errors are distributed normally, this statistic follows
the chi-squared distribution with four degrees of freedom, and the
typical value will be $\chi^2\sim 4$. However the colors of two
quasars at the same redshift, although similar, will in general not be
exactly proportional. Fluctuations about the median color-redshift
relation of quasars \citep{Richards01} will result in an additional
source of `dispersion' in our color similarity statistic.  This leads
to a much broader distribution of $\chi^2$ than expected from Gaussian
statistical errors. Figure~\ref{fig:chi2_dist} shows the distribution
of $\chi^2$ for the 64621 unique pair combinations of 359 quasars in
in the redshift interval $2.4<z<2.45$. The
median value of this distribution is $33.1$, so that a quasar pair
survey which aims to achieve 50\% completeness in this redshift
interval would have to observe all quasar pair candidates with
$\chi^2<33.1$. Also notice that a long tail in this distribution
extends even beyond $\chi^2\sim 100$ because of outliers from the
median color-redshift relation of quasars \citep{Richards01}, caused
by broad absorption line features or reddening \citep{Richards03,Hopk04}.

Our survey for close pairs of quasars thus involves a tradeoff between
completeness and efficiency, since tolerating larger values of
$\chi^2$ will increase the number of false pair candidates. Our follow
up observations targeted quasar pair candidates with $\chi^2<20$. This
threshold implies a certain level of completeness for our survey,
which varies redshift, as the dispersion in the color-redshift
relation of quasars depends on redshift \citep{Richards01}. We
quantify this incompleteness by dividing the SDSS quasar sample into
redshift bins of $\delta z=0.043$, and computing the fraction of all
the unique combinations of pairs in each bin with $\chi^2<20$.
Figure~\ref{fig:chi2_complete} shows the completeness of our quasar
pair survey as a function of redshift. For the redshift range
$0.70<z<3.0$, where most of our binary quasars lie, the $\chi^2<20$
cut results in $\sim 35\%$ completeness.

Finally, note that the statistic defined by eqn.~(\ref{eqn:chi}) uses
an isotropic `metric' in color-space.  This would not be the case had
we included the variance of the color-redshift relation $\sigma^m(z)$
in our errors $\sigma^m_{\rm total}$, similar to the procedure used by
\citet{photoz,Weinstein04} to determine photometric redshifts of
quasars. In retrospect, this would be a more suitable procedure for
finding binary quasars. However, we were also conducting a search for
wide separation gravitational lenses, and for lenses there is no
color-redshift scatter (after all, it is the same quasar observed
twice!). 


We applied our color similarity statistic to all objects within the
annulus $3\arcsec<\Delta\theta<60\arcsec$ around the 59,608 quasars
($0.7 < z < 3.0$) in the 
combined SDSS+2QZ catalog.  We refer to the
binaries selected by searching around the SDSS+2QZ quasars as members
of our `$\chi^2$' sample.  To be considered for follow up
observations, a quasar pair had to meet the following criteria

\bea
\chi^2 &<&20.0\nonumber\\
0.70 < &z& < 3.0\nonumber\\
i&<&21.0 \\
\sigma_i&<&0.2\nonumber.\label{eqn:cuts} 
\eea 

\begin{deluxetable*}{lccccccccc}
\tablecolumns{10}
\tablewidth{0pc}
\tablecaption{Previously Known Binary Quasars in the SDSS Footprint 
\label{table:castles}}

\tablehead{Name              &$z$   & $\Delta\theta$ & $R_{\rm prop}$  &$|\Delta v|$ & $i_1$ & $i_2$ & \phn$\chi^2$ & Sample & Notes}
\startdata 
 SDSSJ~1120$+$6711                 & 1.49 & \phn1.5 & \phn\phn9.0 & $<$100  &  18.49  & 19.57  &   --          &  lens & 1\\
 Q~1120$+$0195                 & 1.47 & \phn6.5 & \phn39.6    &  630   &  15.61  & 20.26  &  \phn20.6     &   --   & 2\\
 HS~1216$+$5032             & 1.46 & \phn9.1 & \phn55.1    &  50    &  16.76  & 18.31  &  534.1        &  overlap & 3\\
 Q~1343$+$2650                   & 2.03 & \phn9.2 & \phn55.4    & 200    &  19.14  & 19.97  &  113.7        &   --   & 4\\
 LBQS~1429$-$008                & 2.08 & \phn5.1 & \phn30.8    &  260   &  17.40  & 20.74  &  \phn\phn4.3  & $\chi^2$ & 5\\
 2QZ~1435$+$0008            & 2.38 &  33.2   &  194.8    & 760    &  19.90  & 20.56  &  \phn\phn7.0  & $\chi^2$ & 6\\
 Q~1635$+$267                   & 1.96 & \phn3.9 & \phn23.2    & \phn30 &  19.04  & 20.26  &  \phn\phn7.7  &  photo & 7 \\
 SDSS~J2336$-$0107      & 1.29 & \phn1.7 & \phn10.0    & 240    &  19.26  & 18.94  &  \phn\phd--   &   lens & 8\\
 Q~2345$+$007 & 2.16 & \phn7.1 & \phn42.4    & 480    &  18.68  & 20.45  &  \phn\phn0.3  & $\chi^2$ & 9\\
 \enddata
\tablecomments{\footnotesize 
The redshift of the binary quasar is $z$, $\Delta
\theta$ is the angular separation,$R$ is the proper transverse
separation, $|\Delta v|$ is the velocity difference between the two
quasars in $\kms$, $i_1$ and $i_2$ are extinction corrected $i$-band magnitudes of the 
brighter
and fainter quasar respectively, and $\chi^2$ is the value of our
color similarity statistic, computed only for pairs with
$\Delta\theta>3\arcsec$.  The last column labeled 'Sample' indicates
which of our selection algorithms recovered the binary.\\ 
$^1$  \citep{Pindor05} SDSS binary quasar\\
$^2$  \citep{MD89} Q~1120+0195 is also named PHL 1222 and UM 144\\ 
$^3$ \citep{Hagen96}. Spectra of both quasars are in the SDSS sample\\ 
$^4$ \citep{Cramp88} Both quasars are below the flux limit of the SDSS quasar survey. \\
$^5$ \citep{Hew89}\\
$^6$ \citep{Miller04}\\
$^7$ \citep{Sram78} This region of sky has been imaged by
the SDSS but not yet spectroscopically observed. Both quasars are
members of the faint photometric quasar sample\\
$^8$ \citep{Gregg02} SDSS binary quasar\\
$^9$ \citep{Weedman82}\\}
\end{deluxetable*}


The minimum redshift was imposed because of our desire to find wide
separation gravitational lenses \citep{Quad03,Quad04}.  The
$\sigma_i<0.2$ requirement gets rid of objects with very large
photometric errors due to problems with deblending or very poor
image quality.  In addition, we required that the companion objects be
optically unresolved in the SDSS imaging to avoid contamination from
galaxies. Nearly all quasars at $z>0.70$ should be unresolved in the
SDSS imaging, with the exception of very small separation pairs
$<3\arcsec$, however, these pair candidates are selected by our 
algorithm described in \S~\ref{sec:lens_selection}. 

We also used the criteria in eqn.~(\ref{eqn:cuts}) to search for pair
candidates in the catalog of 273,287 faint photometric
quasars. We restricted attention to members of the catalog only, 
and did not consider other nearby photometric objects.  We will refer
to binaries discovered in this catalog as our `photometric
sample'. The same criteria as in eqn~(\ref{eqn:cuts}) were used, but
there were no lower or upper limits on redshift since spectroscopic
redshifts are not available for this sample.

\subsection{Overlap and Spectroscopic Selection}

As mentioned previously, the SDSS contains a fair number of
overlapping spectroscopic plates for which fiber collision does not
limit the pair separation.  Furthermore, quasar pairs can be found
below the fiber collision limits by searching the SDSS+2QZ quasar catalog 
over regions where the survey areas overlap. Finally for separations 
$\theta\geq60\arcsec$, larger than the SDSS fiber collision scale
\footnote{Although the fiber collision limit is 55$\arcsec$, for
  operational purposes we take it to be 60$\arcsec$ to give a small
  buffer from the actual limit where the fiber tiling may still be
  imperfect.}, quasar pairs can be found in the entire SDSS area.  We
refer to pairs found in the spectroscopic catalog with
$\theta\leq 60\arcsec$ as our `overlap' sample and we refer to those
with $\theta>60\arcsec$ as our `spectroscopic' sample.  No color
similarity criteria were applied to these objects, and the only
magnitude or error limits are those imposed by the SDSS target
selection \citep{qsoselect}.




\subsection{Previously Known Binaries}

It is instructive to ask whether previously known binary quasars are
selected by our selection techniques.  In Table~\ref{table:castles} we
list the nine previously known binary quasars with $0.7\lesssim z
\lesssim 3.0$ which are in the SDSS footprint, of which seven were
recovered.  The column labeled `sample' indicates the sample for
which each binary was selected as a candidate. Six of these binaries
were listed in the compilation of binary quasars on the
CASTLES\footnote{Available at http://cfa-wwww.harvard.edu/castles.}
website. The others are SDSS~J2336-0107 \citep{Gregg02} and
SDSS~J1120+6711 \citep{Pindor05}, discovered recently in the SDSS
search for gravitational lenses, and 2QZ~J1435+0008, the 33$\arcsec$
pair of quasars discovered in the 2QZ by \citet{Miller04}.  

Of the nine binaries listed in the table, all but Q~1343+2650,
Q~1635+267, and 2QZ~J1435+0008 had at least one member of the pair in
the SDSS spectroscopic sample. Neither member of the pairs Q~1343+2650
and 2QZ~1435+0008 were targeted for spectroscopy because these quasars
are below the flux limit of the SDSS quasar catalog ($i<19.1$ for
quasars in this region of color space). The brighter member of
Q~1635+267 is above the flux limit, but this area of sky has only been
imaged and has yet to be spectroscopically observed. However, both
members of this pair are members of the faint photometric catalog, and
indeed, this pair was selected as a candidate by our photometric
selection.  Both members of the binary HS1216+5032 \citep{Hagen96}
received SDSS fibers, so that this binary is a member of our overlap
sample. The brighter of the two members of the famous double quasar
Q~2345+0007 received an SDSS fiber, and this pair was selected as a
pair candidate by our $\chi^2$ algorithm.  Of the two binaries which
were not recovered, Q~1343+2650 was missed because it was below the
SDSS flux limits and the quasar pair Q~1120+0195 \citep{MD89} was
missed because its $\chi^2=20.6$, is just above the cutoff $\chi^2<
20$ of the `$\chi^2$' selection algorithm.

\section{Spectroscopic Observations}
\label{sec:APO}

Candidates in our lens, $\chi^2$, and photometric samples require follow up 
spectroscopy to confirm the quasar pair hypothesis, which we describe in this 
section. 

The SDSS images of the candidate quasar pairs and the spectrum of the
quasar with an SDSS or 2QZ spectrum were visually inspected to reject
bad imaging data and possible spectroscopic
misidentifications. Color-color diagrams for each candidate were also
visually inspected, and pairs for which the companion object overlaps
the stellar locus \citep[see e.g.][]{Richards01} were given a lower
priority.

The result of a successful follow up observation of a quasar pair
candidate falls into one of four categories: (1) a quasar-quasar pair
at the same redshift (2) a projected pair of quasars at different
redshifts (3) a quasar-star pair (4) a star-star pair (for the
photometric catalog).  As an operational definition, we consider
quasar pairs with velocity differences of $|\Delta v|\leq 2000 \kms$
to be at the same redshift, since this brackets the range of velocity
differences caused by both peculiar velocities, which could be as
large as $\sim 500~\kms$ if binary quasars reside in rich environments,
and redshift uncertainties caused by blueshifted broad lines ($\sim
1500~\kms$) \citep{Richards02}.

Spectra of the vast majority of our quasar pair candidates were
obtained with the Astrophysical Research Consortium (ARC) 3.5m
telescope at the Apache Point Observatory (APO), during a number of
nights between March 2003 and January 2005. In addition, two of the
binary quasars in our sample were confirmed at other telescopes:
SDSSJ1600+0000 was confirmed at the ESO 3.58m New Technology
Telescope, and SDSSJ1028+3929 was discovered at the Hobby-Eberly
Telescope. Higher signal to noise ratio spectra of five of the binary
quasars in our sample were obtained at the 10m Keck I \& Keck II
telescopes.

For the ARC 3.5m observations, we used the Double Imaging Spectrograph
(DIS), a double spectrograph with a transition wavelength of 5350
\AA\ between the blue and red side.  The observations were taken with
low resolution gratings, with a dispersion of 2.4 \AA\ pixel$^{-1}$ in
the blue side and 2.3 \AA\ pixel$^{-1}$ on the red side, and a
resolution of roughly 2 pixels. A $1.5\arcsec$ slit was used and we
oriented the slit at the position angle between the two quasars so
that both members of the pair could be observed simultaneously.  The
final spectrum covers the wavelength range of 3800 \AA\ to 10,000 \AA.
The wavelength scale is calibrated with a polynomial fit to lines from
an Ar-He-Ne lamp; the typical rms error of the fits is smaller than
0.5 \AA.  Observations of a variety of optical spectrophotometric
standards \citep{OG83} provided flux calibration; however, most of the
candidates were not observed under photometric conditions, nor were
they observed with the slit oriented at the parallactic angle.
Exposure times ranged from 1200 seconds for candidates with $i\sim 18$
to 2400 seconds for the faintest candidates, $i\sim 20.8$.


The binary quasar SDSSJ1600+0000 was discovered at the ESO New Technology
Telescope 3.58m on UT 2001 April 18, using the red CCD of
the ESO Multi-Mode Instrument (EMMI). The \#13 grating
(2.66~\AA/pixel, $R \simeq 600$ at 6000~\AA) was used with an OG530
blue-blocking filter. Relative spectrophotometric calibration
was obtained through observations of GD~108 \citep{Oke90}, but 
this calibration is uncertain blueward of 5350~\AA\ and redward of
9300~\AA.

The binary SDSSJ1028+3939 was identified from observations obtained
using the Marcario Low Resolution Spectrograph \citep{HET} on the
Hobby-Eberly Telescope (HET) on UT 2005 January 14. The HET
observations were obtained with a 300 line mm$^{-1}$ grating and GG385
blocking filter.  Spectra of both components were obtained
simultaneously using a 1.5" slit.  The spectra covered the range
4400-8000~\AA\ at a resolution of~18~\AA.


Higher signal to noise ratio spectra of both components of the quasar
pairs SDSSJ0955+6045, SDSSJ1010+0416, and SDSSJ1719+2549 were obtained
on UT 2003 February 5-6, using the Echelle Spectrograph and Imager
\citep[ESI;][]{ESI98} on the Keck II telescope.  The seeing was
$0\farcs 6$. The ESI has a dispersion of $0.15-0.3~$\AA~${\rm
  pixel}^{-1}$ over a wavelength range of $4000-10500~$\AA, and the
$1\arcsec$ slit used for these observations projects to 6.5
pixels. The 900s exposures were obtained with the slit aligned at the
position angle of the components of the pair.

Spectra of both quasars in the pair SDSSJ0248+0009 were obtained on UT
1999 October 17 using the Low-Resolution Imaging Spectrograph Keck II
telescope \citep[LRIS;][]{LRIS}, which was before LRIS was
commissioned as a double spectrograph. The 300 line ${\rm mm}^{-1}$
grating blazed at $5000$~\AA\ was used, giving a spectral coverage of
$5220$~\AA\ and a dispersion of $2.55~$\AA~${\rm pixel}^{-1}$. The
$0\arcsec.7$ longslit was used and the seeing was $0\arcsec.6$.  A
single 900s exposure was obtained with the slit aligned at the
position angle of the components of the pair.

We obtained spectra of both quasars in the pair SDSSJ0048-1051 on UT
2003 September 27 using the LRIS Double Spectrograph on the Keck I
telescope. The D560 dichroic was used, which splits light between the
blue arm and the red arm at $5600$~\AA. On the blue side, the 600 line
${\rm mm}^{-1}$ grism blazed at $4000~$\AA\ was used, giving a spectral
coverage of $2590$~\AA\ and a dispersion of $0.63~$ \AA~${\rm
pixel}^{-1}$. On the red side, the 400 line ${\rm mm}^{-1}$ grating
was used blazed at $8500~$\AA, giving a spectral coverage of
$3810~$\AA\ and a dispersion of $1.86~$\AA~${\rm pixel}^{-1}$. The $1\arcsec.0$
longslit was used and the seeing was $0.\arcsec6$.  A single 900s
exposure was obtained at an airmass of $1.2$ with the slit aligned at
the position angle of the components of the pair.

All the data were reduced using standard procedures in the
IRAF\footnote{IRAF is distributed by the National Optical Astronomy
  Observatories, which are operated by the Association of Universities
  for Research in Astronomy, Inc., under cooperative agreement with
  the National Science Foundation.} package, supplemented by IDL
routines borrowed from the SDSS spectroscopic reduction software and
adapted to the different instruments. Quasar redshifts were determined
by cross correlating the quasar spectra with the first four
eigenspectra of a principal component decomposition of the SDSS quasar
sample \citep{Schlegel05}.

\section{Binary Quasar Sample}
\label{sec:sample}

In this section we present a sample of 218 new quasar pairs with
transverse separations $R_{\rm prop}< 1~\hMpc$ over the redshift
range $0.5 \le z \le 3.0$. Of these, 65 have angular separations
$\theta \le 60\arcsec$, i.e. below the SDSS fiber collision scale. Our 26
new binaries with transverse proper separations $R_{\rm prop} <
50~\hkpc$ ($\theta < 10\arcsec$) more than doubles the number of
known binary quasars with separations this small. Table~\ref{table:binsample} 
lists relevant quantities for 33 binaries with $3\arcsec < \theta \le
60\arcsec$ discovered from our $\chi^2$ and photometric
samples. The last column indicates which algorithm which was used to
select the binary. The binaries with $\theta \le 3\arcsec$ discovered
from our lens sample are shown in Table~\ref{table:lenssample}, and
the overlap and spectroscopic binaries found by searching for
pairs in the SDSS+2QZ quasar catalog are shown in
Table~\ref{table:specsample}. Table~\ref{table:sample} gives a 
summary of the number of binary quasars selected by each algorithm
described in \S~\ref{sec:selection}.  For completeness, the Appendix
includes tables of projected pairs of quasars at different redshifts,
as well as projected quasar-star pairs.  

\begin{figure}
  \centerline{
    \epsfig{file=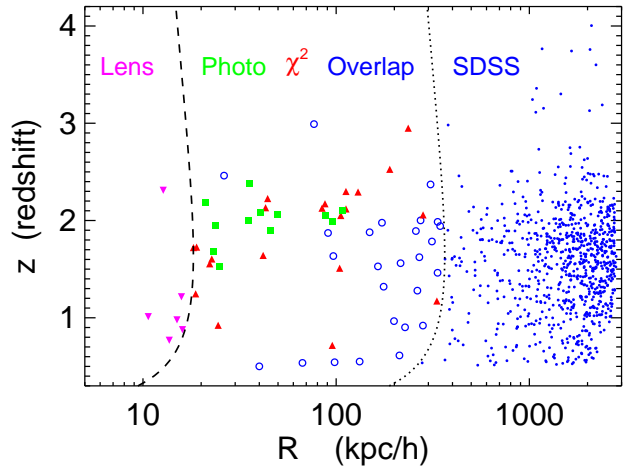,bb=120 72 576 432,width=0.5\textwidth}}
  \caption{Range of redshifts and proper transverse separations probed
    by the binary quasars published in this work (see
    Tables~\ref{table:binsample}, \ref{table:lenssample},
    \ref{table:specsample}, and \ref{table:sample}).  The blue circles
    are binary quasars identified in the SDSS spectroscopic sample.
    The region to the left of the dotted curve is excluded because of
    fiber collisions ($\theta < 60\arcsec$), with the exception of the
    binaries discovered from overlapping plates, which are indicated
    by the larger open blue circles.  The magenta circles are members
    of the lens sample, red circles are from the $\chi^2$ sample, and
    green squares are binaries from the photometric sample. Because
    of the fiber collisions, the vast majority of small separation
    pairs $R\lesssim 300~\hkpc$ were discovered from our follow-up
    observations. The dashed curve indicates the transverse separation
    corresponding to $\theta=3\arcsec$ below which binaries are found
    with our lens algorithm.  Although we publish only pairs with
    separations $R_{\rm prop}<1~\hMpc$ in this work, Figure~\ref{fig:scatter}
    shows all pairs in the SDSS+2QZ catalog out to $3~\hMpc$ for the
    sake of illustration.
    \label{fig:scatter}
  }
\end{figure}

The distribution of redshifts and proper transverse separations probed
by these binary quasars is illustrated by the scatter plot in
Figure~\ref{fig:scatter}. The (magenta) upside down triangle are
members of the lens sample, the (green) squares are members of the
photometric sample, and (red) triangle are in the $\chi^2$ sample. The
(blue) open circles are members of the overlap sample and smaller
(blue) dots are pairs in the spectroscopic sample.  The dashed curve
shows the proper transverse distance corresponding to
$\theta=3\arcsec$, which divides the lens sample from the other
samples, and the dotted line indicates the distance corresponding to
$\theta=60\arcsec$, above the fiber collision limit so that pairs can
be found in the spectroscopic quasar catalog. It should be noted that
the distribution of points in Figure~\ref{fig:scatter} reflects some
biases in our observational program. In particular, we tended to
observe small separation pairs first, and we were much more likely to
observe candidates with $z>2$, because quasar pairs at these redshifts
are of interest for studying the Ly$\alpha$ forest \citep{Hennawi05b}.

We next discuss the possibility that some of the quasar pairs in this
sample are strong gravitational lenses rather than binaries. After
showing spectra of some of the more notable binaries, we define a
sub-sample of binaries selected homogeneously which we will use in our
analysis of small scale quasar clustering in \S~\ref{sec:clustering}.

\subsection{Contamination by Gravitational Lenses}

It is possible that some of the quasar pairs with image splittings
$\lesssim 15^{\prime\prime}$ in our sample could be wide separation
strong gravitational lenses rather than binary quasars. Indeed, the
recently discovered quadruply imaged lensed quasar SDSSJ1004+4112
\citep{Quad03,Quad04} with a maximum image separation of $14\arcsec.6$
was discovered as part of our follow up campaign to discover quasar
pairs, as were two new gravitational lenses with separations $\gtrsim
3\arcsec$ \citep{Oguri05}. We thus review the set of objective
criteria that must be satisfied for a quasar pair to be classified as
a binary or lens \citep{Koch99,MWF99} and briefly discuss why we have
concluded that the binary hypothesis is correct for our pairs.

A quasar pair can be positively confirmed as a binary if the spectra
of the images are vastly different \citep[c.f. ][]{Gregg02}, if only
one of the images is radio-loud \citep[an $O^2R$ pair, in the notation
  of ][]{Koch99}, or if the quasars' hosts are detected and they are
not clearly lensed. The sufficient conditions for a pair to be
identified as a lens are the presence of more than two images in a
lensing configuration, the measurement of a time delay between
images, the detection of a plausible deflector, or the detection of
lensed host galaxy emission.

For the majority of pairs in our sample, the APO discovery spectra
have too low a signal to noise ratio to make convincing arguments for
or against the lensing hypothesis based on spectral dissimilarity.
The exceptions are the five binaries for which we have high signal to
noise ratio Keck spectra (three from ESI and two from LRIS). 
We comment on the spectral similarity of these binaries below.

Although the \emph{absence} of a deflector in images of a quasar pair
does not strictly speaking confirm the binary hypothesis, it certainly
makes it more plausible.  For small separation ($\Delta\theta\lesssim
3\arcsec$)  gravitational lenses, the lens galaxies are rarely detected
in the relatively shallow SDSS imaging. However, for the wider
separation lenses with $\Delta\theta\gtrsim3\arcsec$, like
SDSSJ1004+4112 ($\Delta\theta_{\rm max}=14.62\arcsec$, $z_{\rm
    lens}=0.68$) or Q~0957+561
\citep[$\Delta\theta=6.2\arcsec$, $z_{\rm lens}=0.36$;][]{Walsh79},
where the lens is a bright galaxy in a cluster or group, the lens
galaxies \emph{are} detected in the SDSS imaging, though it is quite
possible that fainter high redshift lens galaxies or clusters in other
wide separation lens systems would go undetected.

Using the University of Hawaii 2.2m telescope, we have taken deeper optical ($i\lesssim 24$; $z\lesssim 22$) or near
infrared images ($J \lesssim 22$; $H\lesssim 22$) of all the binaries
in our sample with separations $\theta\le 4\arcsec$ and of a subset of
those with wider separations.  None of the images showed lens galaxies
in the foreground. For the wider $\theta>4\arcsec$ pairs, the lensing
hypothesis would require a very bright massive galaxy in a group or
cluster. In particular, because the typical redshift range of our
binaries is $z=1.5-2$, the most probable lens redshifts would be in
the range $z=0.3-0.7$, which would likely have been detected in the
SDSS imaging. Furthermore, these wide separation multiply imaged
quasars are extremely rare \citep{Hennawi05a}, thus we are confident
that the pairs in our sample are all binaries.

\subsection{Sample Spectra of Binaries}

\begin{figure*}
  \centerline{
    \epsfig{file=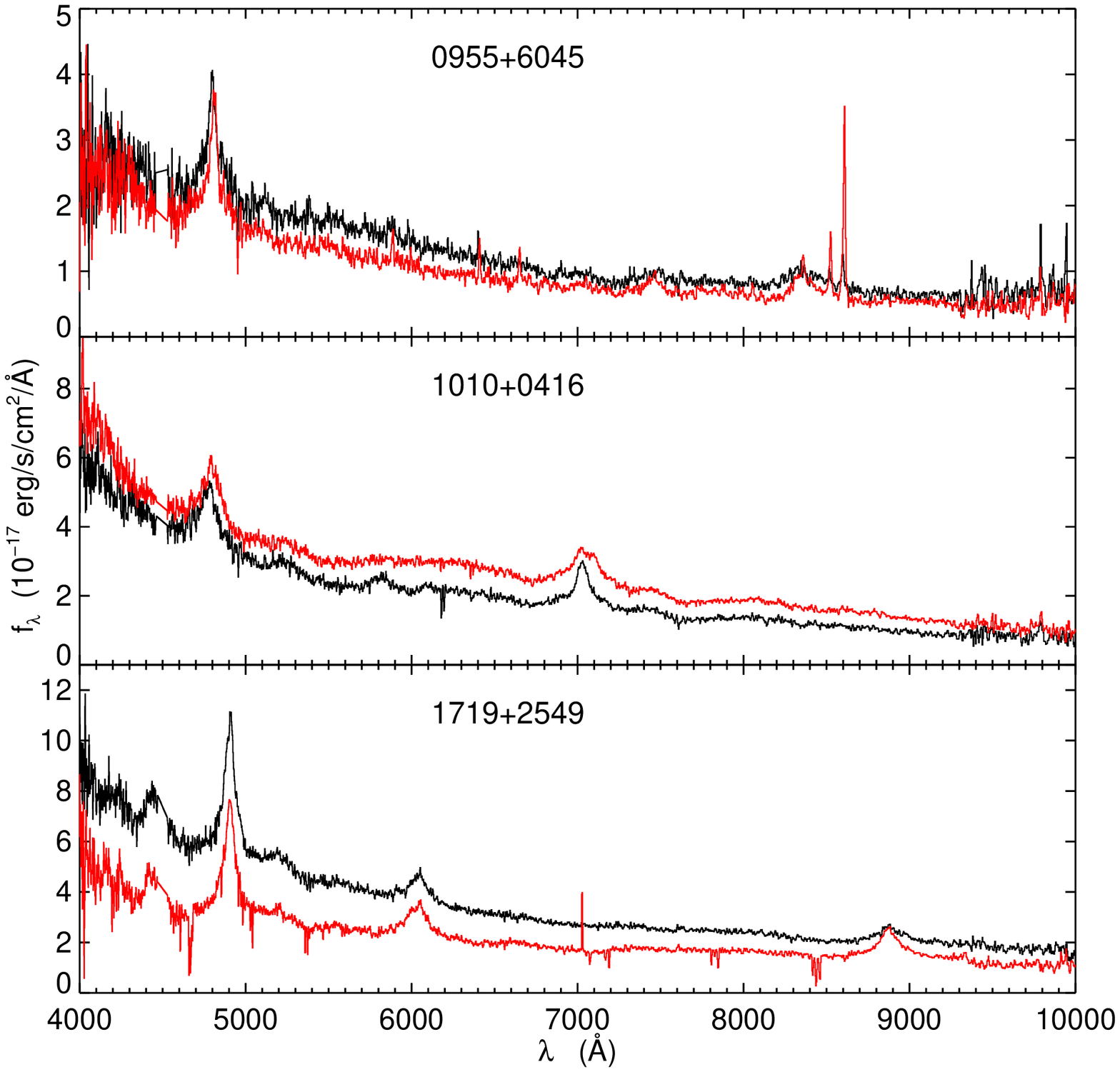,width=1.0\textwidth}}
  \caption{Keck ESI spectra of both members of three binary
    quasars. The top panel is the binary SDSSJ0955+6045 ($z=0.72$,
    $\Delta\theta=18\arcsec.6$, $R_{\rm prop}=95.5~\hkpc$), the middle
    panel is SDSSJ1010+0416 ($z=1.51$, $\Delta\theta=17\arcsec.2$,
    $R_{\rm prop}=104.3~\hkpc$), and the bottom panel is
    SDSSJ1719+2549 ($z=2.17$, $\Delta\theta=14\arcsec.7$, $R_{\rm
      prop}=87.5~\hkpc$). The discontinuity in the spectra at 4500\AA\ 
    is an artifact of a gap in the Echelle orders. 
    \label{fig:esi}}
\end{figure*}

Keck ESI spectra of the three binaries SDSSJ0955+6045, SDSSJ1010+0416,
and SDSSJ1719+2549 are shown in Figure~\ref{fig:esi}. The signal to
noise ratios of these spectra are high enough that we can make
arguments against the lensing hypothesis based on spectral
dissimilarity. In particular, for SDSSJ0955+6045 the narrow [OIII]
emission lines are significantly stronger in one of the quasars
than the other.  In SDSSJ1010+0416, the CIII] emission lines
have a velocity offset of $\sim 2000 \ \kms$, although this offset is
less apparent in MgII, which tends to be a better tracer of the
systemic redshifts of quasars \citep{Richards02}. Finally, for
SDSSJ1719+2549, the peak to continuum flux ratios of the MgII broad
line differ by a factor of $\sim 1.5$ between the two quasars. 

\begin{figure*}
  \centerline{
    \epsfig{file=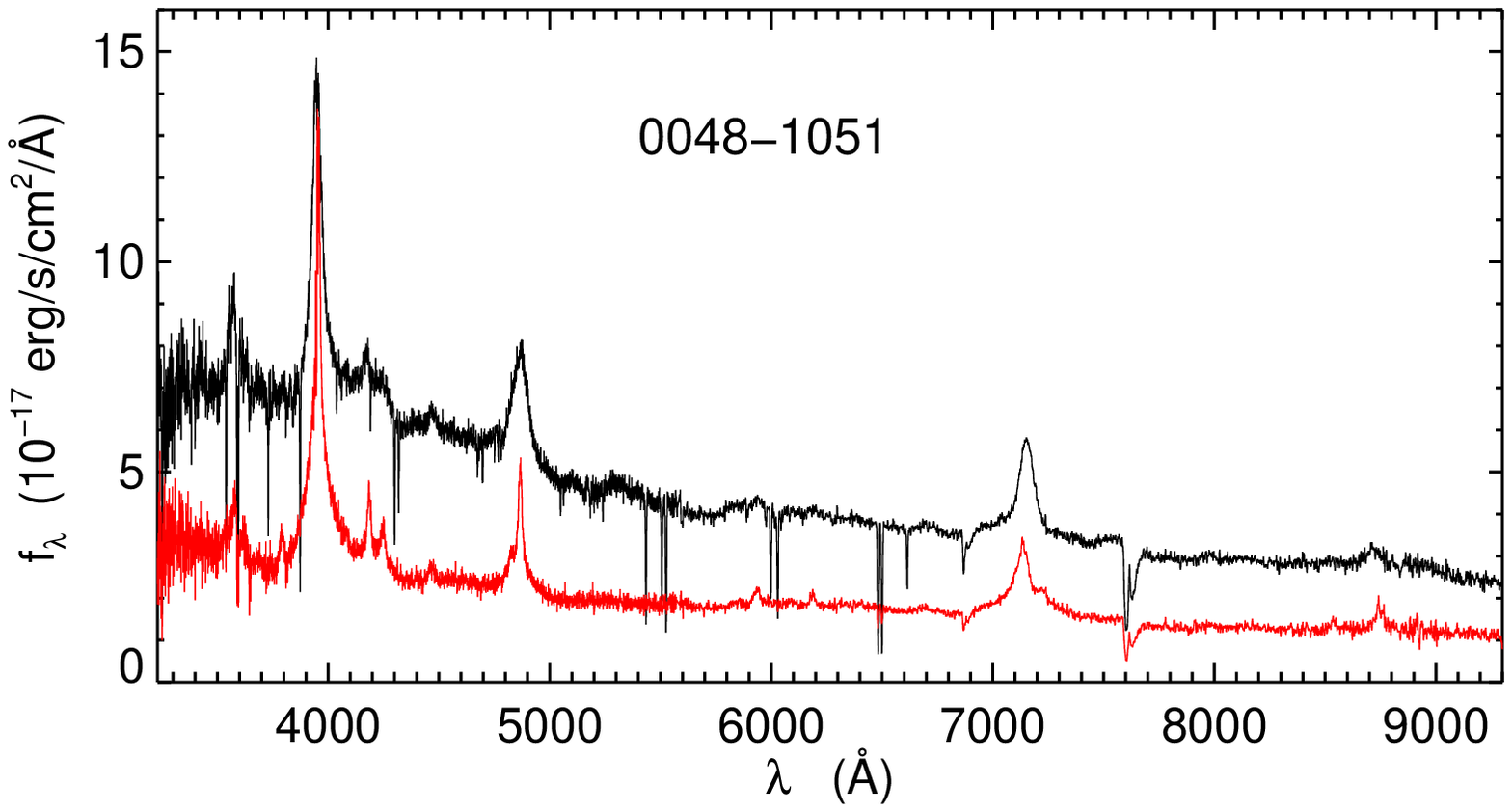,bb=40 0 500 250,width=0.5\textwidth}
    \epsfig{file=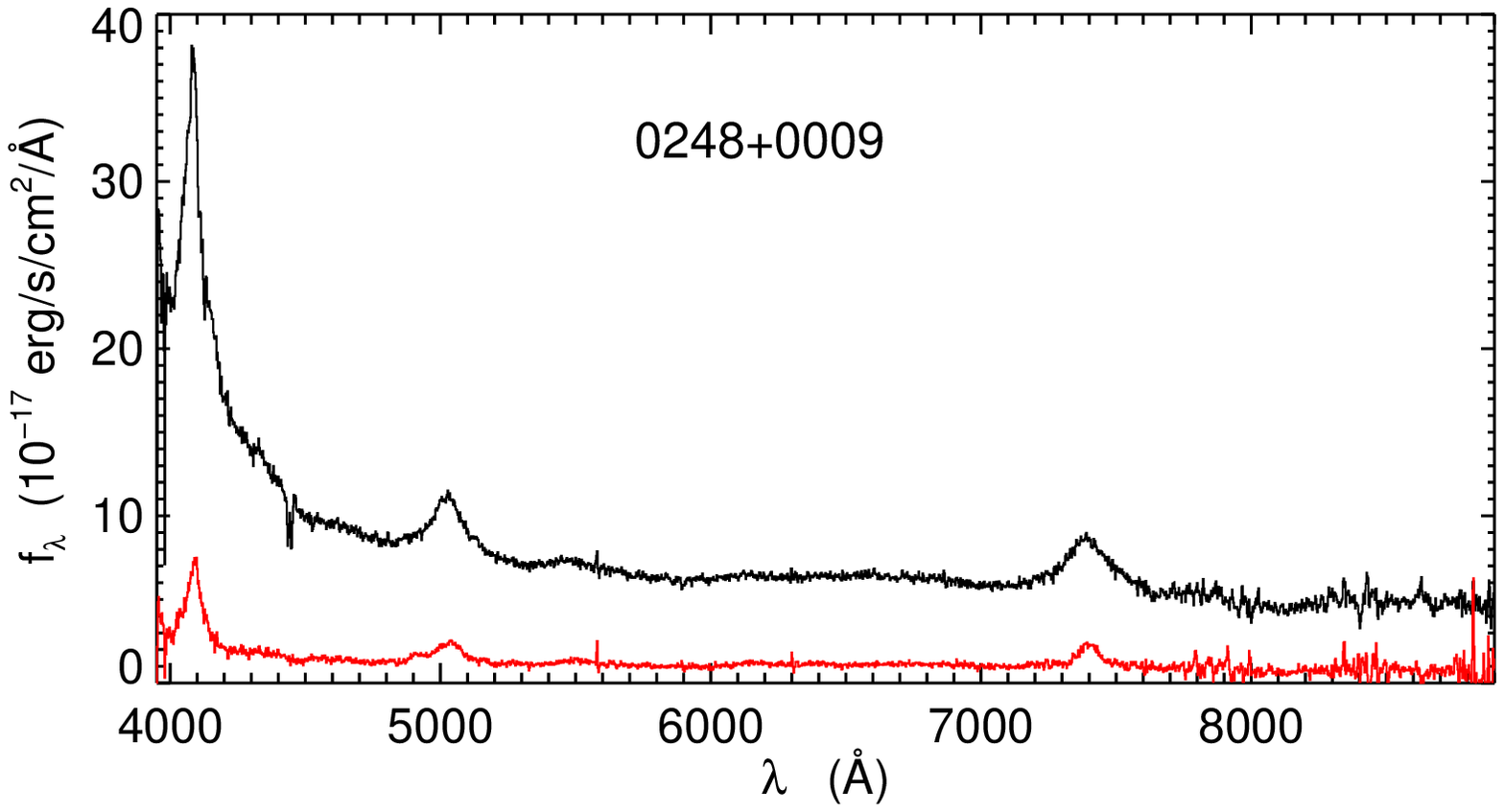,bb=40 0 500 250,width=0.5\textwidth}}
  \caption{Keck spectra of two binary quasars.  Spectral
    dissimilarity and the absence of a lensing galaxy in optical and near IR 
    follow-up imaging provide strong evidence that these are both binary
    quasars rather than gravitational lenses. \emph{Left:} Keck LRIS moderate resolution spectra of both
    members of the binary quasar SDSSJ0048-1051 ($z=1.56$,
    $\Delta\theta=3\arcsec.6$, $R=22.1~\hkpc$). The absorption feature
    at 7600\AA\ is telluric. \emph{Right:} Keck
    LRIS low resolution spectra of both members of SDSSJ0248+0009
    ($z=1.64$, $\Delta\theta=6\arcsec.8$, $R=41.9~\hkpc$).  
    \label{fig:lris}}
\end{figure*}

Figure~\ref{fig:lris} shows Keck LRIS spectra of SDSSJ0048-1051
(moderate resolution) and SDSSJ0248+0009 (low resolution). For
SDSSJ0048-1051 the profiles of all the emission lines differ
significantly, especially MgII. For SDSSJ0248+0009, the peak to
continuum flux ratios differ significantly for CIV, CIII], and MgII.



\begin{figure*}
  \centerline
      {\epsfig{file=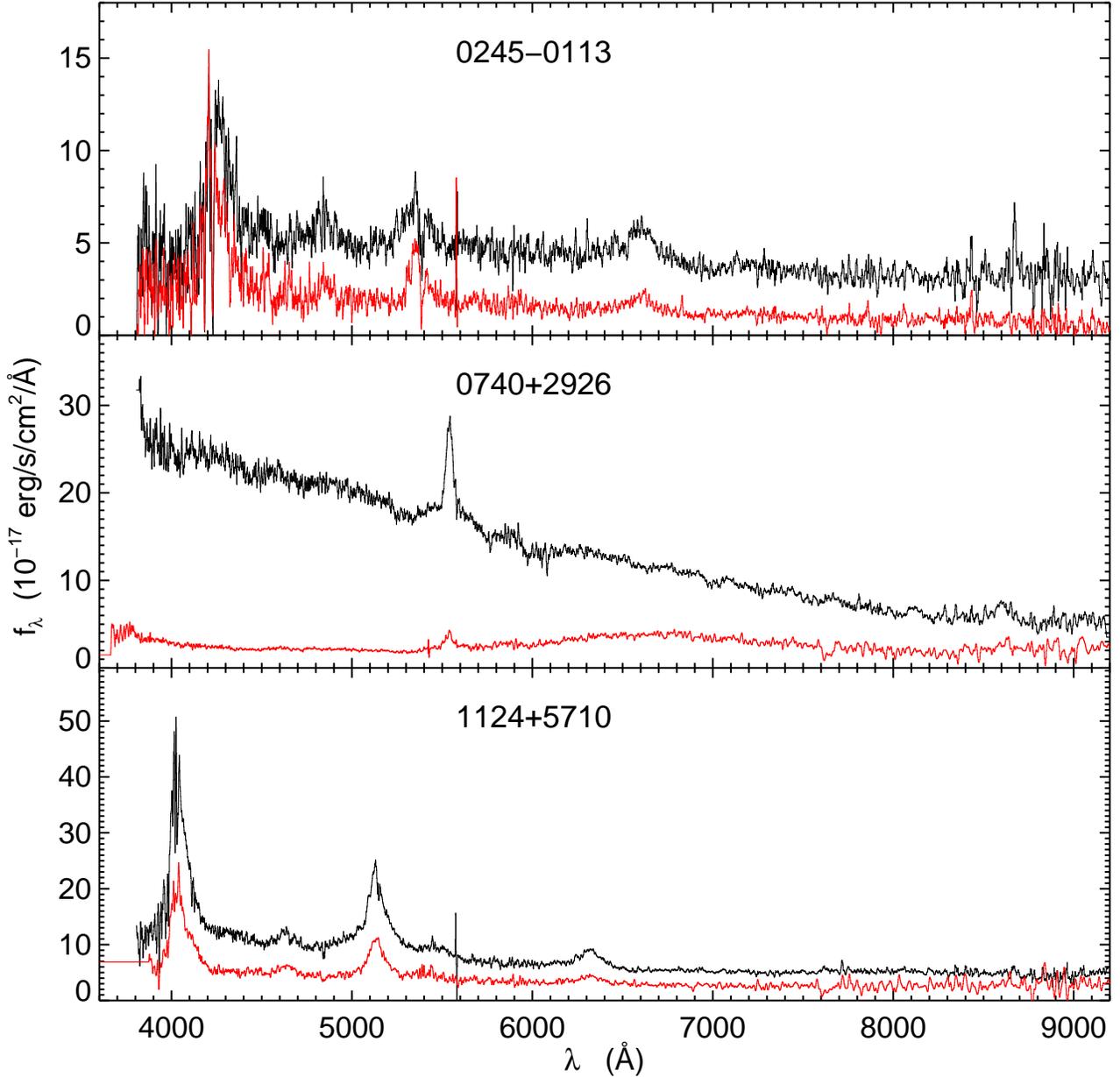,width=1.0\textwidth}}
  \caption{ SDSS and APO spectra of both members of three binary
    quasars.  The top panel is the binary SDSSJ0245-0113 ($z=2.46$,
    $\Delta\theta=4\arcsec.5$, $R_{\rm prop}=26.3~\hkpc$), the middle
    panel is SDSSJ0740+2926 ($z=0.98$, $\Delta\theta=2\arcsec.6$,
    $R_{\rm prop}=15.0~\hkpc$), and the bottom panel is SDSSJ1124+5710
    ($z=2.31$, $\Delta\theta=2\arcsec.2$, $R_{\rm prop}=12.7~\hkpc$).
    The binary in the top panel was a member of our overlap sample,
    hence both quasars have SDSS spectra. For both the middle and
    bottom panels, the black curves are SDSS spectra of the brighter
    quasar in the pair and the red curves are the APO spectra of the
    fainter companions. The absorption feature at 7600\AA \ in the lower
    two spectra is telluric.  Although all three of these binaries have
    separations $\Delta\theta\lesssim 5\arcsec$ characteristic of
    gravitational lenses, deep optical and near IR imaging show no
    lenses in the foreground.
    \label{fig:sdss_apo1}
  }
\end{figure*}
\begin{figure*}
  \centering
  \epsfig{file=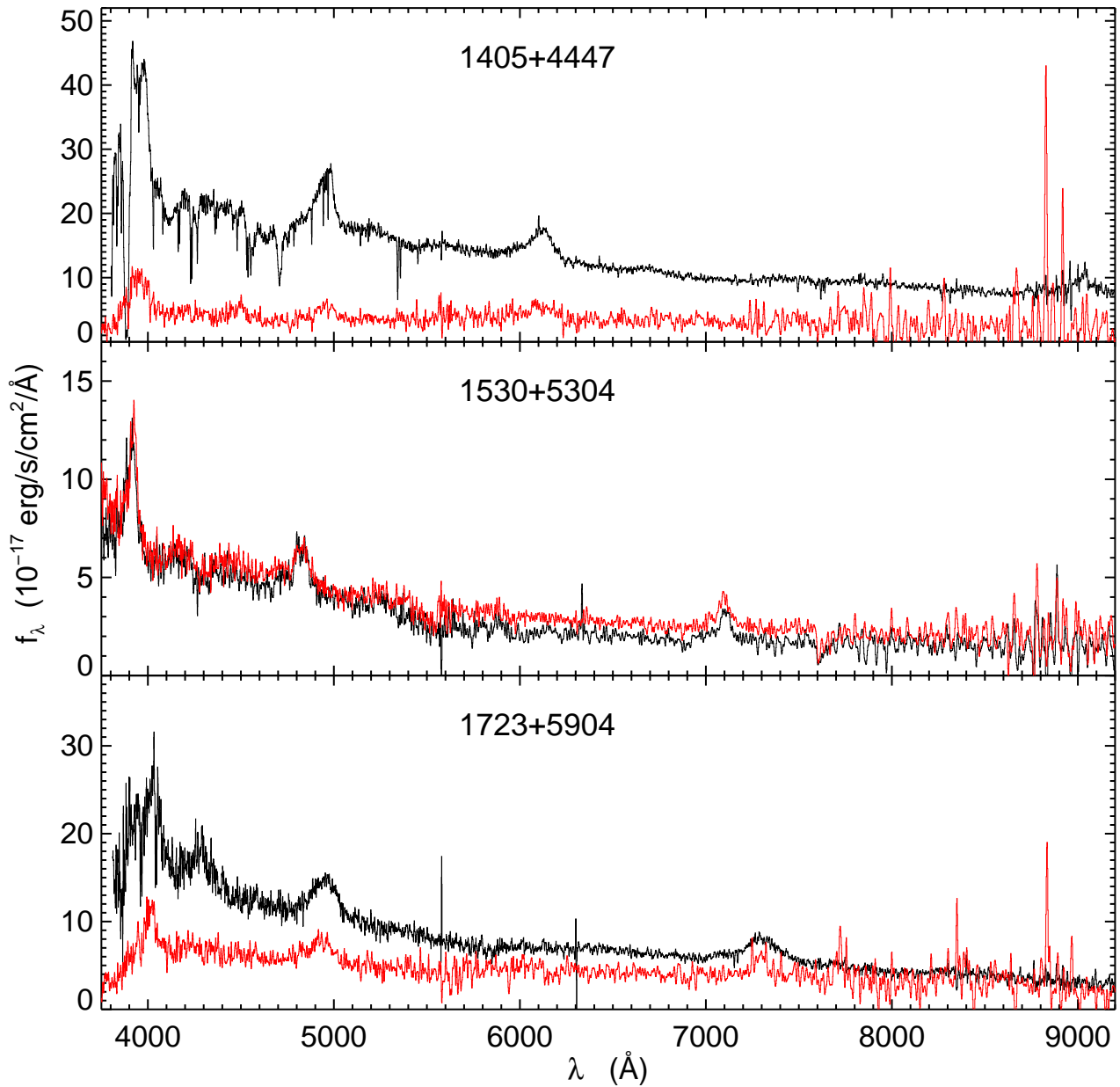,width=1.0\textwidth}
  \caption{ SDSS and APO spectra of both members of three binary
    quasars.  The top panel is the binary SDSSJ1405+44447 ($z=2.23$,
    $\Delta\theta=7\arcsec.4$, $R=142.8~\hkpc$), the middle panel is
    SDSSJ1530+5304 ($z=1.53$, $\Delta\theta=4\arcsec.1$,
    $R=63.3~\hkpc$), and the bottom panel is SDSSJ1723+5904 ($z=1.60$,
    $\Delta\theta=3\arcsec.7$, $R=59.0~\hkpc$).  The binary in the
    middle panel is a member of the photometric sample, and both
    spectra were taken at APO. In the other two panels, the black
    curves are SDSS spectra of the brighter quasar in the pair and the
    red curves are the APO spectra of the fainter companion.  Deep
    imaging of SDSSJ1405+44447 and SDSSJ1723+5904 shows no lens in the
    foreground. The absorption features at 7600\AA\ are telluric.
    \label{fig:sdss_apo2}
  }
\end{figure*}

Figures~\ref{fig:sdss_apo1} and \ref{fig:sdss_apo2} show 
SDSS and APO spectra of six other binaries in our sample. 

\begin{deluxetable*}{rccccccccccccc}
\tablecolumns{14}
\tablewidth{0pc}
\tablecaption{Binary Quasars with separations $3\arcsec < \Delta\theta < 60\arcsec$ Discovered From Follow-Up Observations\label{table:binsample}}
\tablehead{Name & RA (2000) & Dec (2000) &$u$ &$g$ & $r$ & $i$ & $z$ & $\Delta \theta$ & z &$\Delta v$ &  $R_{\rm prop}$  & ~$\chi^2$ & Sample}
\startdata
\hfill SDSSJ0048$-$1051A &   00:48:00.77 &   $-$10:51:48.6 &   20.99  &   20.52  &   20.18  &   19.94  &   19.91  &   \phn3.6  &   1.56  &   $<200$  &   \phn22.1  &   \phn 5.8  & $\chi^2$ \\ 
\hfill APOJ0048$-$1051B  &   00:48:00.96 &   $-$10:51:46.2 &   20.39  &   20.04  &   19.70  &   19.30  &   19.28  &               &             &             &             &             \\ 
\hfill SDSSJ0054$-$0946A &   00:54:08.47 &   $-$09:46:38.3 &   18.17  &   17.90  &   17.71  &   17.51  &   17.31  &    14.1  &   2.13  &   \phn-1600  &   \phn84.5  &    17.5    & $\chi^2$ \\ 
\hfill APOJ0054$-$0946B &   00:54:08.04 &   $-$09:46:25.7 &   20.87  &   20.70  &   20.37  &   20.11  &   19.74  &               &             &             &             &             \\ 
\hfill SDSSJ0201$+$0032A &   02:01:43.49 &   $+$00:32:22.7 &   19.99  &   19.39  &   19.47  &   19.41  &   19.19  &    19.0  &   2.30  &   \phn\phn-520  &    112.4  &   10.8 & $\chi^2$ \\
\hfill APOJ0201$+$0032B &   02:01:42.25 &   $+$00:32:18.5 &   20.80  &   20.30  &   20.14  &   20.12  &   19.92  &               &             &             &             &             \\ 
\hfill SDSSJ0248$+$0009A &   02:48:20.80 &   $+$00:09:56.7 &   19.44  &   19.24  &   19.23  &   18.98  &   19.00  &   \phn6.9  &   1.64  &   $<200$  &   \phn41.9  &    \phn8.8  & $\chi^2$ \\ 
\hfill APOJ0248$+$0009B &   02:48:21.26 &   $+$00:09:57.3 &   20.77  &   20.71  &   20.74  &   20.57  &   20.39  &               &             &             &             &             \\ 
\hfill APOJ0332$-$0722A &   03:32:38.38 &   $-$07:22:15.9 &   20.24  &   20.29  &   20.00  &   19.78  &   19.63  &    18.1  &   2.10  &   \phn\phn960  &    108.4  &    \phn5.0  & photo    \\ 
\hfill APOJ0332$-$0722B &   03:32:37.19 &   $-$07:22:19.6 &   20.63  &   20.57  &   20.22  &   19.96  &   19.70  &               &             &             &             &            \\ 
\hfill SDSSJ0846$+$2749A &   08:46:31.77 &   $+$27:49:21.9 &   19.55  &   19.57  &   19.66  &   19.49  &   19.26  &    18.8  &   2.12  &   \phn\phn-490  &    112.4  &   \phn14.3  & $\chi^2$     \\ 
\hfill APOJ0846$+$2749B &   08:46:30.38 &   $+$27:49:18.1 &   19.82  &   19.88  &   19.82  &   19.71  &   19.55  &               &             &             &             &             \\ 
\hfill SDSSJ0939$+$5953A &   09:39:48.78 &   $+$59:53:48.7 &   20.40  &   19.85  &   19.80  &   19.79  &   19.45  &    32.6  &   2.53  &   \phn\phn290  &    189.4  &    \phn9.3  & $\chi^2$      \\ 
\hfill APOJ0939$+$5953B &   09:39:46.56 &   $+$59:53:20.7 &   19.31  &   18.67  &   18.57  &   18.56  &   18.42  &               &             &             &             &             \\ 
\hfill SDSSJ0955$+$6045A &   09:55:24.37 &   $+$60:45:51.0 &   20.84  &   20.37  &   20.29  &   20.25  &   20.34  &    18.6  &   0.72  &   \phn\phn460  &   \phn95.5  &    10.8 & $\chi^2$  \\
\hfill APOJ0955$+$6045B &   09:55:25.37 &   $+$60:45:33.8 &   20.68  &   20.65  &   20.62  &   20.72  &   20.30  &               &             &             &             &             \\ 
\hfill APOJ0959$+$5449A &   09:59:07.46 &   $+$54:49:06.7 &   20.26  &   20.06  &   19.95  &   19.76  &   19.73  &   \phn3.9  &   1.95  &   \phn\phn200  &   \phn23.8  &  18.9  & photo     \\ 
\hfill APOJ0959$+$5449B &   09:59:07.05 &   $+$54:49:08.4 &   20.48  &   20.61  &   20.43  &   20.28  &   19.82  &               &             &             &             &             \\ 
\hfill SDSSJ1010$+$0416A &   10:10:04.98 &   $+$04:16:36.2 &   20.18  &   20.08  &   20.00  &   19.87  &   20.08  &    17.2  &   1.51  &     $<200$  &    104.3  &      \phn6.5  & $\chi^2$  \\ 
\hfill APOJ1010$+$0416B &   10:10:04.37 &   $+$04:16:21.6 &   20.29  &   20.19  &   20.03  &   19.87  &   19.94  &               &             &             &             &             \\ 
\hfill SDSSJ1014$+$0920A &   10:14:11.43 &   $+$09:20:47.7 &   19.95  &   19.14  &   19.12  &   18.92  &   18.73  &    22.0  &   2.29  &     $<200$  &    129.8  &        11.7   & $\chi^2$    \\ 
\hfill APOJ1014$+$0921B &   10:14:10.29 &   $+$09:21:01.7 &   20.81  &   20.27  &   20.25  &   20.13  &   19.79  &               &             &             &             &             \\ 
\hfill HETJ1028$+$3929A &   10:28:43.67 &   $+$39:29:36.9 &   19.96  &   19.90  &   19.95  &   19.57  &   19.49  &   \phn7.5  &   1.89  &   \phn-1030  &   \phn45.6  &    \phn6.4 & photo    \\ 
\hfill HETJ1028$+$3929B &   10:28:44.30 &   $+$39:29:34.8 &   20.92  &   20.81  &   20.91  &   20.60  &   20.66  &               &             &             &             &            \\
\hfill SDSSJ1034$+$0701A &  10:34:51.47 &   +07:01:21.2   &   19.86  &   19.47  &   18.94  &   19.04  &   19.08  &    3.1  &   1.25  &   \phn\phn850  &    18.7  &        \phn9.7   & $\chi^2$      \\ 
\hfill APOJ1034$+$0701B &   10:34:51.38 &   +07:01:24.0   &   21.20  &   21.26  &   21.08  &   20.89  &   20.99  &               &             &             &             &             \\ 
\hfill 2QZJ1056$-$0059A &   10:56:44.89 &   $-$00:59:33.4 &   20.16  &   19.92  &   19.89  &   19.80  &   19.59  &   \phn7.2  &   2.13  &   \phn\phn-350  &   \phn43.1  &   \phn6.7 & $\chi^2$ \\ 
\hfill APOJ1056$-$0059B &   10:56:45.25 &   $-$00:59:38.1 &   20.76  &   20.78  &   20.59  &   20.58  &   20.33  &               &             &             &             &             \\ 
\hfill SDSSJ1123$+$0037A &   11:23:10.96 &   $+$00:37:45.2 &   19.02  &   19.03  &   18.90  &   18.98  &   19.08  &    56.3  &   1.17  &   \phn\phn-330  &    332.1  &      \phn6.0  & $\chi^2$  \\ 
\hfill APOJ1123$+$0037B &   11:23:07.21 &   $+$00:37:45.7 &   20.26  &   20.29  &   20.10  &   20.12  &   20.22  &               &             &             &             &             \\ 
\hfill APOJ1225$+$5644A &   12:25:45.73 &   $+$56:44:40.7 &   20.05  &   19.28  &   19.44  &   19.35  &   19.07  &   \phn6.0  &   2.38  &   \phn\phn970  &   \phn35.4  &    20.7   & photo   \\ 
\hfill APOJ1225$+$5644B &   12:25:45.24 &   $+$56:44:45.1 &   21.08  &   20.52  &   20.50  &   20.35  &   19.80  &               &             &             &             &             \\ 
\hfill SDSSJ1254$+$6104A &   12:54:21.98 &   $+$61:04:22.0 &   19.24  &   19.06  &   19.01  &   18.92  &   18.74  &    17.6  &   2.05  &   \phn-1010  &    105.6  &          11.4  & $\chi^2$         \\ 
\hfill APOJ1254$+$6104B &   12:54:20.52 &   $+$61:04:36.0 &   19.67  &   19.56  &   19.47  &   19.28  &   19.14  &               &             &             &             &             \\ 
\hfill APOJ1259$+$1241A &   12:59:55.62 &   $+$12:41:53.8 &   20.33  &   19.99  &   19.93  &   19.74  &   19.53  &   \phn3.6  &   2.19  &   \phn\phn-840  &   \phn21.2  &   \phn6.0 & photo \\ 
\hfill APOJ1259$+$1241B &   12:59:55.46 &   $+$12:41:51.0 &   20.22  &   19.90  &   19.87  &   19.79  &   19.56  &               &             &             &             &             \\ 
\hfill APOJ1303$+$5100A &   13:03:26.17 &   $+$51:00:47.5 &   20.46  &   20.33  &   20.28  &   20.05  &   20.03  &   \phn3.8  &   1.68  &   \phn\phn220  &   \phn23.0  &   \phn4.0 & photo  \\ 
\hfill APOJ1303$+$5100B &   13:03:26.13 &   $+$51:00:51.3 &   20.87  &   20.60  &   20.66  &   20.38  &   20.59  &               &             &             &             &             \\ 
\hfill SDSSJ1310$+$6208A &   13:10:37.89 &   $+$62:08:21.6 &   18.87  &   18.77  &   18.63  &   18.57  &   18.35  &    46.9  &   2.06  &   \phn-1850  &    281.7  &    10.8   & $\chi^2$        \\ 
\hfill APOJ1310$+$6208B &   13:10:31.96 &   $+$62:08:43.5 &   20.65  &   20.49  &   20.35  &   20.15  &   20.12  &               &             &             &             &             \\ 
\hfill SDSSJ1337$+$6012A &   13:37:13.13 &   $+$60:12:06.7 &   18.68  &   18.57  &   18.56  &   18.34  &   18.42  &   \phn3.1  &   1.73  &   \phn\phn-610  &   \phn18.9  &   \phn 0.6 & $\chi^2$ \\ 
\hfill APOJ1337$+$6012B &   13:37:13.08 &   $+$60:12:09.8 &   20.15  &   20.01  &   20.03  &   19.66  &   19.80  &               &             &             &             &             \\ 
\hfill SDSSJ1349$+$1227A &   13:49:29.84 &   $+$12:27:07.0 &   17.92  &   17.76  &   17.71  &   17.46  &   17.45  &   \phn3.0  &   1.72  &    $<200$  &   \phn18.3  &    12.9 & $\chi^2$ \\ 
\hfill APOJ1349$+$1227B &   13:49:30.00 &   $+$12:27:08.8 &   19.50  &   19.26  &   19.10  &   18.74  &   18.60  &               &             &             &             &             \\ 
\hfill APOJ1400$+$1232A &   14:00:52.07 &   $+$12:32:35.2 &   20.41  &   20.28  &   20.27  &   20.13  &   19.88  &    14.6  &   2.05  &   \phn1470  &   \phn87.7  &   \phn 0.6  & photo  \\ 
\hfill APOJ1400$+$1232B &   14:00:52.56 &   $+$12:32:48.0 &   20.54  &   20.42  &   20.45  &   20.27  &   19.99  &               &             &             &             &             \\ 
\hfill SDSSJ1405$+$4447A &   14:05:01.94 &   $+$44:47:59.9 &   18.68  &   18.14  &   17.93  &   17.86  &   17.66  &   \phn7.4  &   2.23  &   \phn1870  &   \phn44.2  &   \phn 3.2  & $\chi^2$   \\ 
\hfill APOJ1405$+$4447B &   14:05:02.41 &   $+$44:47:54.4 &   20.61  &   20.12  &   19.96  &   19.89  &   19.60  &               &             &             &             &             \\ 
\hfill APOJ1409$+$3919A &   14:09:53.74 &   $+$39:19:60.0 &   20.31  &   20.24  &   20.14  &   20.06  &   19.78  &   \phn6.8  &   2.08  &   \phn\phn480  &   \phn40.7  &   \phn3.9 & photo  \\ 
\hfill APOJ1409$+$3919B &   14:09:53.88 &   $+$39:19:53.4 &   20.89  &   20.92  &   20.87  &   20.67  &   20.31  &               &             &             &             &             \\ 
\hfill APOJ1530$+$5304A &   15:30:38.56 &   $+$53:04:04.2 &   20.86  &   20.54  &   20.42  &   20.13  &   20.06  &   \phn4.1  &   1.53  &   \phn\phn230  &   \phn25.0  &   \phn8.8 & photo  \\ 
\hfill APOJ1530$+$5304B &   15:30:38.82 &   $+$53:04:00.8 &   20.67  &   20.66  &   20.53  &   20.31  &   20.21  &               &             &             &             &             \\ 
\hfill SDSSJ1546$+$5134A &   15:46:10.55 &   $+$51:34:29.5 &   22.34  &   20.57  &   20.31  &   20.20  &   20.29  &    42.2  &   2.95  &   \phn-1450  &    236.0  &      12.4 & $\chi^2$ \\ 
\hfill APOJ1546$+$5134B &   15:46:14.24 &   $+$51:34:05.0 &   20.64  &   19.43  &   19.10  &   18.91  &   18.88  &               &             &             &             &             \\ 
\hfill SDSSJ1629$+$3724A &   16:29:02.59 &   $+$37:24:30.8 &   19.47  &   19.26  &   19.10  &   19.16  &   19.06  &   \phn4.4  &   0.92  &    $<200$  &   \phn24.5  &    15.9 & $\chi^2$ \\ 
\hfill APOJ1629$+$3724B &   16:29:02.63 &   $+$37:24:35.2 &   19.50  &   19.41  &   19.28  &   19.38  &   19.31  &               &             &             &             &             \\ 
\hfill SDSSJ1719$+$2549A &   17:19:46.66 &   $+$25:49:41.2 &   20.17  &   19.93  &   19.90  &   19.85  &   19.61  &    14.7  &   2.17  &   \phn\phn-220  &   \phn87.5  &   \phn 6.4  & $\chi^2$\\ 
\hfill APOJ1719$+$2549B &   17:19:45.87 &   $+$25:49:51.3 &   20.08  &   19.74  &   19.65  &   19.65  &   19.46  &               &             &             &             &             \\ 
\hfill SDSSJ1723$+$5904A &   17:23:17.42 &   $+$59:04:46.8 &   19.40  &   19.04  &   18.95  &   18.69  &   18.76  &   \phn3.7  &   1.60  &   \phn\phn-830  &   \phn22.7  &   \phn3.0 & $\chi^2$ \\ 
\hfill APOJ1723$+$5904B &   17:23:17.30 &   $+$59:04:43.2 &   21.24  &   20.43  &   20.46  &   20.18  &   20.20  &               &             &             &             &             \\ 
\hfill APOJ2128$-$0617A &   21:28:57.38 &   $-$06:17:50.9 &   19.84  &   19.66  &   19.87  &   19.80  &   19.68  &   \phn8.3  &   2.07  &   \phn\phn-290  &   \phn49.7  &    15.7  & photo  \\ 
\hfill APOJ2128$-$0617B &   21:28:57.74 &   $-$06:17:57.2 &   20.03  &   19.74  &   20.12  &   19.92  &   19.63  &               &             &             &             &            \\ 
\hfill APOJ2214$+$1326A &   22:14:27.03 &   $+$13:26:57.0 &   20.39  &   20.19  &   20.19  &   19.96  &   19.60  &   \phn5.8  &   2.00  &   \phn\phn-690  &   \phn35.2  &   19.4 & photo  \\ 
\hfill APOJ2214$+$1326B &   22:14:26.79 &   $+$13:26:52.3 &   20.65  &   20.41  &   20.26  &   19.98  &   19.82  &               &             &             &             &           \\ 
\hfill APOJ2220$+$1247A &   22:20:30.26 &   $+$12:47:33.5 &   20.11  &   20.03  &   20.00  &   19.88  &   19.78  &    15.9  &   1.99  &   \phn1600  &   \phn95.5  &     43.0    & photo     \\ 
\hfill APOJ2220$+$1247B &   22:20:29.53 &   $+$12:47:45.1 &   20.92  &   20.85  &   20.72  &   20.34  &   20.21  &               &             &             &             &           \\ 
\enddata \tablecomments{\footnotesize Quasars labeled SDSS or 2QZ are
  members of the SDSS or 2QZ spectroscopic quasar catalog and are
  designated as quasar `A'. Quasars discovered from follow up
  spectroscopy are labeled APO (or HET) and designated `B'. For pairs
  discovered from the photometric catalog, both quasars are labeled
  APO (or HET) and `A' designates the brighter of the two
  quasars. Extinction corrected SDSS five band PSF photometry are
  given in the columns $u$, $g$, $r$, $i$, and $z$.  The redshift of
  quasar `A' is indicated by column z, $\Delta \theta$ is the angular
  separation in arcseconds, $\Delta v$ is the is the velocity of
  quasar B relative to quasar A in $\kms$, $R_{\rm prop}$ is the
  transverse proper separation in $\hkpc$, and $\chi^2$ is the value
  of our color similarity statistic. The last column labeled 'Sample'
  indicates the selection algorithm used to find the binary.\\ }
\end{deluxetable*}


\begin{deluxetable*}{rccccccccccc}
\tablecolumns{12}
\tablewidth{0pc}
\tablecaption{Binary Quasars with $\Delta \theta < 3\arcsec$ Discovered from Lens Selection\label{table:lenssample}}
\tablehead{Name & RA (2000) & Dec (2000) &$u$ &$g$ & $r$ & $i$ & $z$ & $\Delta \theta$ & z &$\Delta v$ &  $R_{\rm prop}$}
\startdata
\hfill SDSSJ0740$+$2926A & 07:40:13.45 & $+$29:26:48.4 & 18.61 & 18.46 & 18.30 & 18.42 & 18.48 & 2.6 & 0.98 &  230  & 15.0\\
\hfill APOJ0740$+$2926B  & 07:40:13.43 & $+$29:26:45.7 & 19.98 & 19.67 & 19.50 & 19.68 & 20.03 &            &       &     \\ 
\hfill SDSSJ1035$+$0752A & 10:35:19.37 & $+$07:52:58.0 & 19.17 & 19.13 & 18.97 & 19.03 & 19.14 & 2.7 & 1.22 &  270  & 15.8\\
\hfill APOJ1035$+$0752B  & 10:35:19.23 & $+$07:52:56.4 & 20.62 & 20.42 & 19.98 & 19.84 & 19.84 &            &       &     \\
\hfill SDSSJ1124$+$5710A & 11:24:55.24 & $+$57:10:57.0 & 19.34 & 18.66 & 18.83 & 18.65 & 18.44 & 2.2 & 2.31 & -540  & 12.7\\
\hfill APOJ1124$+$5710B  & 11:24:55.44 & $+$57:10:58.4 & 20.31 & 19.83 & 19.52 & 19.52 & 19.42 &            &       &     \\
\hfill SDSSJ1138$+$6807A & 11:38:09.21 & $+$68:07:38.8 & 18.28 & 17.98 & 17.87 & 17.89 & 17.78 & 2.6 & 0.77 &  840  & 13.7\\
\hfill APOJ1138$+$6807B  & 11:38:08.89 & $+$68:07:36.9 & 20.31 & 19.74 & 19.76 & 19.72 & 19.58 &            &       &     \\ 
\hfill SDSSJ1508$+$3328A & 15:08:42.21 & $+$33:28:02.6 & 17.88 & 17.78 & 17.81 & 17.97 & 17.86 & 2.9 & 0.88 & $<$200 & 16.0\\
\hfill APOJ1508$+$3328B  & 15:08:42.22 & $+$33:28:05.5 & 20.56 & 20.36 & 20.15 & 20.56 & 19.71 &     &      &       &      \\
\hfill SDSSJ1600$+$0000A$^{\dagger}$ & 16:00:15.50 & $+$00:00:45.5 & 19.23 & 19.11 & 18.84 & 18.95 & 19.08 & 1.9 & 1.01 & -660  & 10.6 \\
\hfill NTTJ1600$+$0000B$^{\dagger}$ & 16:00:15.59 & $+$00:00:46.9 &  --   &   --  & $\approx$ 21 & $\approx$ 21 & --  &   &       &   \\
\enddata

\tablecomments{\footnotesize Quasars labeled SDSS are the members of
  the SDSS spectroscopic quasar catalog and are designated as quasar
  `A'. Quasars discovered from follow up spectroscopy are labeled APO
  (or NTT) and designated `B'. Extinction corrected SDSS five band PSF
  photometry are given in the columns $u$, $g$, $r$, $i$, and $z$.
  These magnitudes are estimated from the deblending algorithm of the
  main SDSS photometric pipeline \citep{Stoughton02} (except for
  1600+0000 see below).  The redshift of the SDSS quasar is given by
  $z$, $\Delta \theta$ is the angular separation in arcseconds,
  $\Delta v$ is the the velocity of quasar B relative to quasar A in
  $\kms$,and $R_{\rm prop}$ is the transverse proper separation in
  $\hkpc$. \\ 
  $^{\dagger}$ The pair SDSSJ1600$+$0000A and
  NTTJ1600$+$0000B was not deblended by the photometric pipeline
  because the separation is too small. The magnitudes of
  SDSSJ1600$+$0000A have contributions from both members of the pair
  and the approximate $r$ and $i$ band magnitudes of NTTJ1600$+$0000B,
  were measured from follow up NTT imaging of this system.}
\end{deluxetable*}


\begin{deluxetable*}{rcccccccccccc}
\tablecolumns{13}
\tablewidth{0pc}
\tablecaption{Binary Quasars Discovered in Overlapping Plates and the SDSS+2QZ Catalog$^{\dagger}$\label{table:specsample}}
\tablehead{Name & RA (2000) & Dec (2000) &$u$ &$g$ & $r$ & $i$ & $z$ & $\Delta \theta$ & z &$\Delta v$ &  $R_{\rm prop}$  & $\chi^2$}
\startdata
\hfill SDSSJ0012$+$0052A &   00:12:01.88 &   $+$00:52:59.7 &   21.53  &   20.82  &   20.33  &   19.83  &   19.63  &    16.0  &   1.63  &   \phn1590  &   \phn97.3  &   \phn59.8  \\ 
\hfill SDSSJ0012$+$0053B &   00:12:02.35 &   $+$00:53:14.1 &   21.03  &   20.82  &   20.51  &   20.22  &   20.10  &               &             &             &             &     \\ 
\hfill SDSSJ0117$+$0020A &   01:17:58.84 &   $+$00:20:21.5 &   17.96  &   17.67  &   17.82  &   17.79  &   17.99  &    44.5  &   0.61  &   \phn\phn-260  &    212.9  &   \phn34.1 \\ 
\hfill SDSSJ0117$+$0021B &   01:17:58.00 &   $+$00:21:04.1 &   20.26  &   20.01  &   20.13  &   19.86  &   19.89  &               &             &             &             &      \\ 
\hfill SDSSJ0141$+$0031A &   01:41:11.63 &   $+$00:31:45.9 &   20.23  &   20.19  &   20.11  &   19.87  &   19.96  &    42.9  &   1.89  &   \phn1140  &    259.2  &   \phn\phn2.9   \\ 
\hfill SDSSJ0141$+$0031B &   01:41:10.41 &   $+$00:31:07.1 &   20.76  &   20.59  &   20.50  &   20.29  &   20.20  &               &             &             &             &      \\ 
\hfill SDSSJ0245$-$0113A &   02:45:12.08 &   $-$01:13:14.0 &   20.56  &   19.85  &   19.56  &   19.47  &   19.33  &   \phn4.5  &   2.46  &   \phn\phn-190  &   \phn26.3  &   \phn17.0 \\ 
\hfill SDSSJ0245$-$0113B &   02:45:11.90 &   $-$01:13:17.6 &   21.14  &   20.57  &   20.45  &   20.38  &   20.11  &               &             &             &             &     \\ 
\hfill SDSSJ0258$-$0003A &   02:58:15.55 &   $-$00:03:34.2 &   18.93  &   18.93  &   18.66  &   18.66  &   18.74  &    29.4  &   1.32  &   \phn\phn240  &    176.6  &   \phn32.9  \\ 
\hfill SDSSJ0258$-$0003B &   02:58:13.66 &   $-$00:03:26.5 &   19.67  &   19.89  &   19.66  &   19.75  &   19.71  &               &             &             &             &     \\ 
\hfill SDSSJ0259$+$0048A$^{\ast}$ &   02:59:59.69 &   $+$00:48:13.7 &   19.63  &   19.26  &   19.22  &   19.33  &   19.10  &    19.6  &   0.89  &    \phn\phn830  &    108.1  &    1914.2
\\
\hfill SDSSJ0300$+$0048B  &   03:00:00.57 &   $+$00:48:28.0 &   19.47  &   19.01  &   16.51  &   16.37  &   16.05  &               &             &             &             & \\

\hfill SDSSJ0350$-$0031A &   03:50:53.29 &   $-$00:31:14.7 &   20.35  &   20.17  &   19.45  &   18.99  &   18.62  &    45.5  &   2.00  &   \phn\phn-920  &    273.8  &    492.9   \\ 
\hfill SDSSJ0350$-$0032B &   03:50:53.05 &   $-$00:32:00.1 &   19.66  &   19.40  &   19.32  &   19.29  &   19.16  &               &             &             &             &     \\ 
\hfill SDSSJ0743$+$2054A &   07:43:37.29 &   $+$20:54:37.1 &   20.06  &   19.86  &   19.77  &   19.51  &   19.43  &    35.5  &   1.56  &   \phn\phn640  &    215.5  &   \phn15.9   \\ 
\hfill SDSSJ0743$+$2055B &   07:43:36.85 &   $+$20:55:12.1 &   20.36  &   20.08  &   20.04  &   19.90  &   19.94  &               &             &             &             &      \\ 
\hfill SDSSJ0747$+$4318A &   07:47:59.02 &   $+$43:18:05.4 &   19.52  &   19.24  &   19.21  &   18.89  &   18.75  &   \phn9.2  &   0.50  &   \phn\phn150  &   \phn39.9  &   \phn\phn6.3\\ 
\hfill SDSSJ0747$+$4318B &   07:47:59.66 &   $+$43:18:11.5 &   19.79  &   19.45  &   19.36  &   19.11  &   18.99  &               &             &             &             &          \\ 
\hfill SDSSJ0824$+$2357A &   08:24:40.61 &   $+$23:57:09.9 &   18.72  &   18.51  &   18.69  &   18.58  &   18.59  &    14.9  &   0.54  &   \phn\phn-170  &   \phn67.0  &   \phn13.2  \\ 
\hfill SDSSJ0824$+$2357B &   08:24:39.83 &   $+$23:57:20.3 &   19.00  &   18.67  &   18.88  &   18.70  &   18.72  &               &             &             &             &        \\ 
\hfill SDSSJ0856$+$5111A &   08:56:25.63 &   $+$51:11:37.4 &   18.90  &   18.52  &   18.64  &   18.45  &   18.51  &    21.8  &   0.54  &   \phn\phn\phn60  &   \phn98.2  &    138.9   \\ 
\hfill SDSSJ0856$+$5111B &   08:56:26.71 &   $+$51:11:18.2 &   20.03  &   19.55  &   19.42  &   19.17  &   19.19  &               &             &             &             &      \\ 
\hfill SDSSJ0909$+$0002A &   09:09:24.01 &   $+$00:02:11.0 &   16.65  &   16.68  &   16.61  &   16.39  &   16.34  &    15.0  &   1.87  &   \phn1700  &   \phn90.6  &   \phn28.0 \\ 
\hfill SDSSJ0909$+$0002B &   09:09:23.13 &   $+$00:02:04.0 &   20.08  &   20.06  &   20.11  &   19.97  &   19.82  &               &             &             &             &    \\ 
\hfill SDSSJ0955$+$0616A &   09:55:56.38 &   $+$06:16:42.5 &   17.79  &   18.07  &   17.84  &   17.81  &   17.86  &    44.0  &   1.28  &   \phn-1040  &    263.3  &   \phn38.6 \\ 
\hfill SDSSJ0955$+$0617B &   09:55:59.03 &   $+$06:17:01.9 &   20.35  &   20.29  &   20.11  &   20.21  &   20.29  &               &             &             &             &   \\ 
\hfill SDSSJ1032$+$0140A &   10:32:44.65 &   $+$01:40:20.5 &   18.93  &   18.86  &   18.84  &   18.76  &   18.86  &    55.1  &   1.46  &   \phn-1580  &    333.5  &   \phn16.0  \\ 
\hfill 2QZJ1032$+$0139B &   10:32:43.17 &   $+$01:39:30.0 &   20.50  &   20.46  &   20.27  &   20.15  &   20.21  &               &             &             &             &    \\ 
\hfill SDSSJ1103$+$0318A &   11:03:57.72 &   $+$03:18:08.3 &   18.35  &   18.36  &   18.30  &   18.10  &   17.96  &    57.3  &   1.94  &   \phn-1730  &    345.7  &   \phn41.5 \\ 
\hfill SDSSJ1104$+$0318B &   11:04:01.49 &   $+$03:18:17.5 &   19.08  &   19.04  &   19.12  &   19.02  &   18.97  &               &             &             &             &   \\ 
\hfill SDSSJ1107$+$0033A &   11:07:25.70 &   $+$00:33:53.9 &   18.98  &   18.94  &   18.84  &   18.51  &   18.42  &    24.8  &   1.88  &   \phn\phn280  &    150.1  &   \phn22.9 \\ 
\hfill 2QZJ1107$+$0034B &   11:07:27.08 &   $+$00:34:07.6 &   20.01  &   20.04  &   20.05  &   19.84  &   19.80  &               &             &             &             &     \\ 
\hfill SDSSJ1116$+$4118A &   11:16:11.74 &   $+$41:18:21.5 &   20.35  &   18.53  &   18.16  &   17.94  &   17.96  &    13.8  &   2.99  &   \phn\phn890  &   \phn76.8  &   \phn28.4  \\ 
\hfill SDSSJ1116$+$4118B &   11:16:10.69 &   $+$41:18:14.4 &   21.33  &   19.44  &   19.17  &   19.03  &   19.03  &               &             &             &             &       \\ 
\hfill SDSSJ1134$+$0849A &   11:34:57.74 &   $+$08:49:35.3 &   19.30  &   19.10  &   19.01  &   18.83  &   18.85  &    27.1  &   1.53  &   \phn\phn-390  &    164.4  &   \phn\phn6.4 \\ 
\hfill SDSSJ1134$+$0849B &   11:34:59.38 &   $+$08:49:23.3 &   19.65  &   19.50  &   19.34  &   19.11  &   19.13  &               &             &             &             &        \\ 
\hfill 2QZJ1146$-$0124A &   11:46:52.97 &   $-$01:24:46.4 &   20.54  &   20.23  &   20.17  &   19.93  &   19.65  &    28.5  &   1.98  &   \phn\phn-490  &    172.0  &   \phn26.0 \\ 
\hfill 2QZJ1146$-$0124B &   11:46:51.19 &   $-$01:24:56.3 &   20.53  &   20.46  &   20.52  &   20.35  &   20.24  &               &             &             &             &      \\ 
\hfill SDSSJ1152$-$0030A &   11:52:40.53 &   $-$00:30:04.3 &   18.95  &   18.80  &   18.93  &   18.86  &   18.80  &    29.3  &   0.55  &   \phn\phn740  &    132.8  &   \phn23.1 \\ 
\hfill 2QZJ1152$-$0030B &   11:52:40.10 &   $-$00:30:32.9 &   20.32  &   20.12  &   20.16  &   19.99  &   20.16  &               &             &             &             &     \\ 
\hfill SDSSJ1207$+$0115A &   12:07:00.97 &   $+$01:15:39.4 &   19.03  &   18.94  &   18.84  &   18.87  &   18.85  &    35.4  &   0.97  &   \phn\phn-260  &    200.1  &   \phn\phn3.5 \\ 
\hfill 2QZJ1207$+$0115B &   12:07:01.40 &   $+$01:15:04.7 &   20.62  &   20.40  &   20.30  &   20.37  &   20.36  &               &             &             &             &         \\ 
\hfill 2QZJ1217$+$0006A &   12:17:36.18 &   $+$00:06:57.4 &   19.96  &   19.93  &   20.06  &   19.76  &   19.62  &    51.8  &   1.78  &   \phn\phn-200  &    314.0  &   \phn13.5  \\ 
\hfill 2QZJ1217$+$0006B &   12:17:35.03 &   $+$00:06:08.6 &   19.17  &   19.30  &   19.32  &   19.05  &   19.17  &               &             &             &             &     \\ 
\hfill 2QZJ1217$+$0055A &   12:17:36.95 &   $+$00:55:22.7 &   20.33  &   20.24  &   20.12  &   20.23  &   19.84  &    40.5  &   0.90  &   \phn\phn\phn-60  &    224.7  &   \phn18.3 \\ 
\hfill 2QZJ1217$+$0055B &   12:17:34.25 &   $+$00:55:22.2 &   20.25  &   19.92  &   19.78  &   19.82  &   19.98  &               &             &             &             &         \\ 
\hfill SDSSJ1226$-$0112A &   12:26:24.09 &   $-$01:12:34.5 &   17.47  &   17.39  &   17.23  &   17.27  &   17.28  &    50.6  &   0.92  &   \phn\phn100  &    282.4  &   \phn22.8 \\ 
\hfill 2QZJ1226$-$0113B &   12:26:25.58 &   $-$01:13:19.9 &   19.74  &   19.82  &   19.75  &   19.79  &   19.53  &               &             &             &             &      \\ 
\hfill SDSSJ1300$-$0156A &   13:00:45.56 &   $-$01:56:31.8 &   18.26  &   18.26  &   18.14  &   17.88  &   17.90  &    44.5  &   1.62  &   \phn\phn480  &    270.7  &   \phn37.6 \\ 
\hfill 2QZJ1300$-$0157B &   13:00:44.52 &   $-$01:57:13.5 &   19.81  &   19.72  &   19.73  &   19.59  &   19.70  &               &             &             &             &     \\ 
\hfill 2QZJ1328$-$0157A &   13:28:30.14 &   $-$01:57:32.8 &   19.86  &   19.49  &   19.52  &   19.55  &   19.42  &    52.6  &   2.37  &   \phn\phn-890  &    309.2  &   \phn26.8 \\ 
\hfill 2QZJ1328$-$0157B &   13:28:33.64 &   $-$01:57:27.9 &   20.46  &   19.92  &   19.78  &   19.77  &   19.75  &               &             &             &             &     \\ 
\hfill 2QZJ1354$-$0108A &   13:54:40.40 &   $-$01:08:45.4 &   19.48  &   19.54  &   19.46  &   19.29  &   19.21  &    55.5  &   1.99  &   \phn\phn740  &    334.2  &   \phn37.1 \\ 
\hfill 2QZJ1354$-$0107B &   13:54:39.97 &   $-$01:07:50.3 &   20.40  &   20.13  &   19.99  &   19.85  &   19.64  &               &             &             &             &     \\ 
\enddata 
\tablecomments{\footnotesize Quasars labeled SDSS or 2QZ are
  members of the SDSS or 2QZ spectroscopic quasar catalog. The
  brighter of the two quasars is designated `A', except for SDSS-2QZ
  pairs for which the SDSS quasar is designated `A'.  
  Extinction corrected SDSS five band PSF photometry are
  given in the columns $u$, $g$, $r$, $i$, and $z$.  The redshift of
  quasar `A' is indicated by column $z$, $\Delta \theta$ is the
  angular separation in arcseconds, $\Delta v$ is the velocity
  of quasar B relative to quasar A in $\kms$, $R_{\rm prop}$ is the transverse
  proper separation in $\hkpc$, and $\chi^2$ is the value of our color similarity
  statistic.\\
  $^{\dagger}$ We publish all quasars with proper transverse separations 
  $R_{\rm prop}<1~\hMpc$, however only those with $\theta < 60\arcsec$ are
  listed in this table. The entire sample of pairs is published
  in the electronic version of this article. \\
  $^{\ast}$ The quasar SDSSJ0300$+$0048B has a large BAL trough which explains the very large 
  $\chi^2=1914.2$
  for this pair.\\ }

\end{deluxetable*}


\begin{deluxetable}{lc}
\tablecolumns{2}
\tablewidth{0pc}
\tablecaption{Summary of Binary Quasar Sample\label{table:sample}}
\tablehead{Algorithm & Number of Binaries}
\startdata
Lens           & 6 \\
$\chi^2$       & 21\\
Photometric    & 12\\
Overlap        & 26\\
Spectroscopic  & 153\\

\enddata
\tablecomments{Number of binary quasars with $R_{\rm prop} < 1~\hMpc$, selected by
the various algorithms discussed in \S~\ref{sec:selection} and plotted
in Figure~\ref{fig:scatter}.}
\end{deluxetable}

\subsection{Clustering Sub-sample}

\begin{figure}
  \centerline{
    \epsfig{file=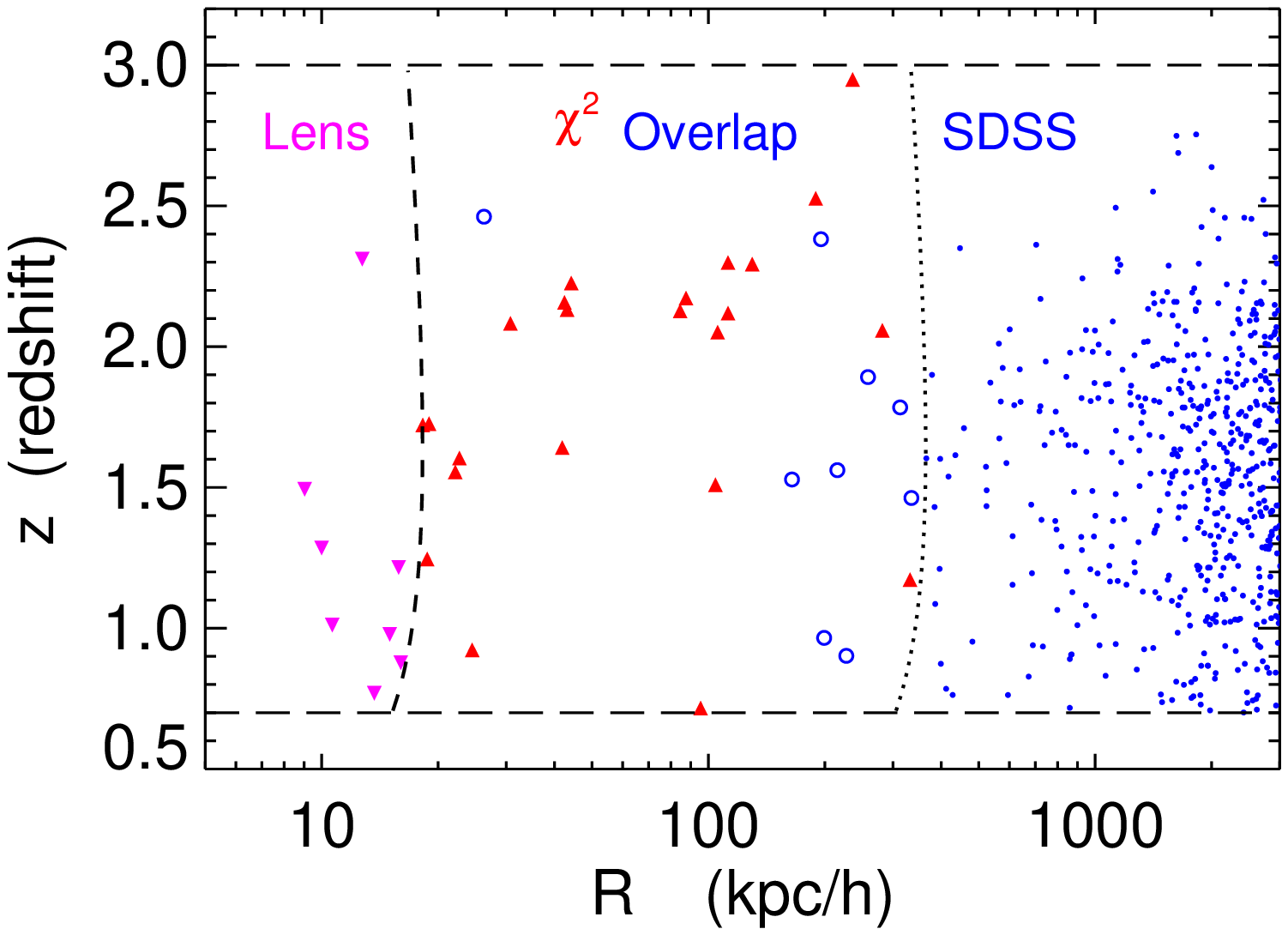,bb=120 72 576 432,width=0.5\textwidth}}
  \caption{Range of redshifts and proper transverse separations probed
    by the binary quasars in our clustering sub-sample.  The blue
    circles are binary quasars identified in the SDSS spectroscopic
    sample ($\theta > 60\arcsec$). The short dashed line indicates the
    transverse separation corresponding to $\theta=60\arcsec$. The
    open blue circles indicate pairs from the overlapping plates.
    These pairs are required to also meet the $\chi^2$ selection
    criteria (eqn.~\ref{eqn:cuts}).  The magenta circles are members
    of the lens sample and red circles are from $\chi^2$ sample.
    The dashed curve indicates the transverse separation
    corresponding to $\theta=3\arcsec$ below which binaries are found
    with our lens algorithm. The horizontal long dashed curves at $z=0.7$ 
    and $z=3.0$ indicate the redshift limits of the clustering sub-sample.
    \label{fig:scatter_cluster}
  }
\end{figure}


In this section, we define a statistical sub-sample of binary quasars which we 
will use to measure the quasar correlation function in \S
\ref{sec:clustering}.  In \S \ref{sec:selection}, the various samples
used to identify the binary quasars in Tables~\ref{table:binsample}, 
\ref{table:lenssample}, and ~\ref{table:specsample} were
described. These various selection algorithms selected quasar pairs over 
different angular scales, with different limiting magnitudes, and varying 
degrees of completeness. Here we combine these samples in a coherent way,
which will allow us to quantify their selection function. 
We pay special attention to the completeness of each sample used and
the parent sample of quasars searched to define each sample.

For the clustering analysis we restrict attention to quasars in the
redshift range $0.7 \leq z \leq 3.0$ with velocity differences
$|\Delta v|<2000 \kms$. We use the lens, $\chi^2$,
overlap, and spectroscopic samples. Our approach is to stitch the samples 
together as a function of angle. The photometric quasar pairs 
are not in the SDSS+2QZ catalog so the selection function 
and completeness of these binaries is more difficult to quantify. 



All pairs with $\theta \leq 3\arcsec$ come from the lens sample. The
parent sample of quasars for this angular range is the 39,142 SDSS
quasars to which we applied the lens algorithm. The completeness is
the product of the \emph{intrinsic} completeness of the lens algorithm
\citep{Pindor03, Inada03} and the fraction of candidates which have
had spectroscopic confirmation thus far in the survey.

For pairs in the range $3\arcsec < \theta \leq 60\arcsec$ we use the
$\chi^2$ sample.  Quasar pairs from the overlap sample which
\emph{also} meet the $\chi^2$ selection criteria in
eqn.~(\ref{eqn:cuts}) are included, and can be thought of as follow up
observations which came for `free' from the overlapping plates.  Of
the 21 binaries with $0.7 \leq z \leq 3.0$ in the overlap sample, 8
satisfy $\chi^2<20$ (see Table~\ref{table:specsample}), and 
are included in the clustering sub-sample. The
completeness of binary quasars in the range $3\arcsec < \theta \leq
60\arcsec$ is the product of completeness of the $\chi^2$ selection,
$C(z|\chi^2<20)$, and the fraction of candidates observed thus
far. The parent sample around which we searched with the $\chi^2$
algorithm is the combined SDSS+2QZ quasar sample of 59,608 quasars
in the range $0.7 \leq z \leq 3.0$.

For separations $\theta > 60\arcsec$ , we use pairs found
in the SDSS spectroscopic catalog of 52,279 quasars. We restrict
attention to the SDSS (rather than SDSS+2QZ), because the completeness
for detecting quasar companions is very high if we restrict attention
to companions above the SDSS flux limit for low redshift
quasars. \citet{vanden05} measured a completeness of $\sim 95\%$ for
quasars in the range $0.3 < z < 3.0$ with $i<19.1$.  Thus we only
include quasar pairs $\theta > 60\arcsec$ in our clustering sample
provided that \emph{at least one} member of the pair is brighter than
this flux limit.

Finally, any of the previously known binaries listed in
Table~\ref{table:castles} which satisfied any of the criteria for 
the clustering sub-sample are also included. Thus we include the binaries
SDSS~J1120+6711 and SDSS~J2336-0107 as part of the lens sample, 
and LBQS~1429-0008, Q~2345+007, and 2QZ~1435+0008 are included 
as part of our $\chi^2$ sample. 

Of the 65 quasar pairs with angular separations $\theta \le 60\arcsec$ 
which we publish in this work, 35 are included in our clustering sub-sample
along with five previously known binaries for a total of 40 sub-arcminute
pairs. The distribution of redshifts and proper transverse separations of 
our clustering sample is is illustrated by the scatter plot in
Figure~\ref{fig:scatter_cluster}. The horizontal long-dashed lines indicate
the redshift limits $0.7\le z \le 3.0$ of the sample, and the symbols and 
dotted and short dashed curves are the same as in Figure~\ref{fig:scatter}.





\section{Clustering Analysis}
\label{sec:clustering}

\begin{figure}
  \centering
  \epsfig{file=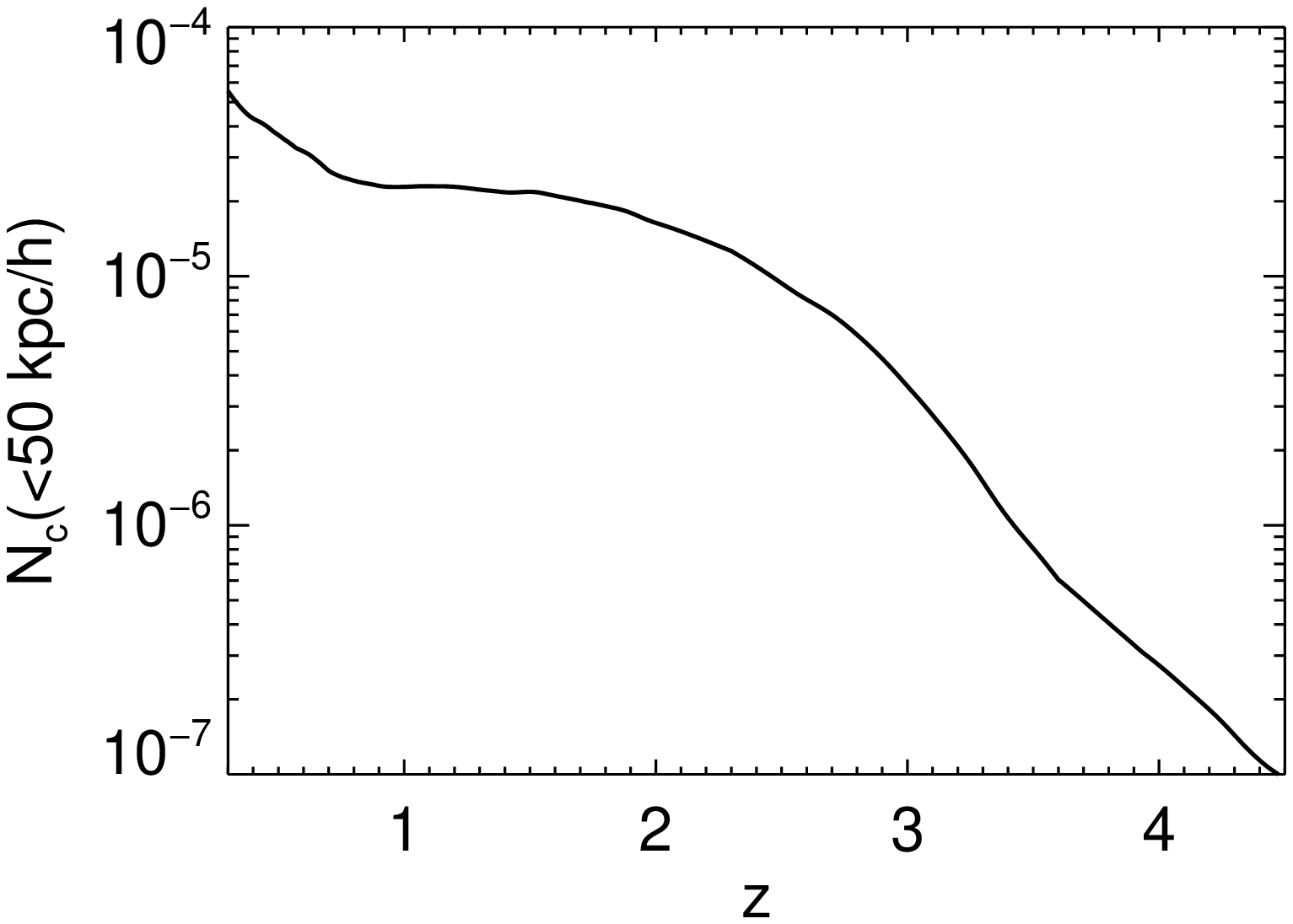,bb=100 72 576 432,width=0.5\textwidth}
  \caption{ Expected mean number of companions per quasar with proper
    transverse separation $R_{\rm prop}<50~\hkpc$ and velocity difference
    $|\Delta v|<2000~\kms$, as a function of redshift calculated from
    eqn.~\ref{eqn:Nc}, which assumes the large scale quasar correlation
    function can be extrapolated as a power law to small scales. We used the 
    correlation function parameters, $\gamma=1.53$ and
    $r_0=4.8~\hMpc$, measured by PMN from the 2dF quasar survey. A
    perfect survey, with no sources of incompleteness, is assumed.  In a
    sample of $\sim 50,000$ quasars, the predicted number of binaries with
    $R<50~\hkpc$ is $\sim 1$. However, our clustering sample already
    contains 20 quasars with transverse separations this small, which
    is compelling evidence for excess clustering over what is expected
    from extrapolating the correlation function.
    \label{fig:Nc}}
\end{figure}

A measurement of quasar clustering on the small scales probed by our
binary sample $10~\hkpc \lesssim R_{\rm prop} \lesssim 1~\hMpc$ is
unprecedented.  Our strategy has been to overcome the fiber collision 
limitation and high level of shot noise by following up close pairs
of quasars to magnitudes $i<21$ fainter than the flux limit of the SDSS
quasar survey.



As our pairs have a fainter flux limit than the underlying parent
quasar catalog, and the mean density at this limit cannot be determined
from the brighter sample, we cannot use the conventional technique of
Monte Carlo integration of a random catalog to compute the correlation
function. However, the mean number density of quasars as a function of
magnitude and redshift is well known \citep{Croom03,Richards05} and
can be computed from the quasar luminosity function. Below, we explain
how we model the quasar luminosity function. We then introduce an
estimator for the quasar correlation function which takes the
incompleteness of our pair survey into account. After determining the
selection function of our clustering sub-sample, we compute its
correlation function and compare it to previous clustering
measurements extrapolated down to the scales $R_{\rm prop}\lesssim
1~\hMpc$ probed by our binaries.

\subsection{Modeling the Luminosity Function}

At low redshift $z<2.3$, the quasar luminosity function has been measured by
several groups \citep{Boyle00, Croom04,Richards05}. We use the 
double power law B-band luminosity function \citep{Boyle00}
\be
\Phi(M_{\rm B},z)=\frac{\Phi_{\ast}}{10^{0.4(\beta_{l}+1)[M_{\rm B}-M^{\ast}_{\rm B}(z)]} + 10^{0.4(\beta_{h}+1)[M_{\rm B}-M^{\ast}_{\rm B}(z)]}},
\label{eqn:Boyle}
\ee
where $\beta_l=-1.64$, $\beta_h=-3.43$, and 
$\Phi_{\ast}=360(h\slash 0.50)^3$ Gpc$^{-3}$mag$^{-1}$. 
The evolution of the break  luminosity $M^{\ast}_{\rm B}(z)$ follows
\be
M^\ast_{\rm B}=M^{\ast}_B(0)-2.5(k_1z+k_2z^2),
\label{eqn:mstar}
\ee
with $k_1=1.36$, $k_2=-0.27$, and $M^{\ast}_B(0)=-21.15+5\log h$. 

The quasar luminosity function is poorly constrained between redshifts
$2.6\leq z \leq 2.9$, because quasar colors cross the stellar locus
\citep[see e.g.][]{Richards01}, in this range and color selected
samples suffer high incompleteness.  As we desire to predict the
number density of quasars in the redshift range $0.7 < z < 3.0$ probed
by our clustering sub-sample, we devise a simple interpolation 
scheme to cover the range $2.3 < z < 3.0$: 
\citet{WLlum02} used a simple analytical model to fit the double power
law luminosity function in eqn. (\ref{eqn:Boyle}) to both the
\citet{Fan01} high redshift ($z>3.6$) luminosity function and the
\citet{Boyle00} low redshift ($z<2.3$) luminosity function.  For
redshifts $z<2.3$ we use the \citet{Boyle00} expression in
eqn.~(\ref{eqn:Boyle}). In the range $2.3 < z < 3.0$ we simply
linearly interpolate between eqn.~\ref{eqn:Boyle} and the \citet{WLlum02} fit.
Since the number of 
binary quasars in our clustering sample in this redshift range is 
relatively small, our conclusions will be insensitive 
to any uncertainties in this procedure. 

Although the luminosity functions we are considering are expressed in
terms of $M_{\rm B}$, both the SDSS spectroscopic survey and our binary quasar
survey have flux limits in the $i$-band.  Thus in order to
compute the number of quasars above our flux limit, we need to know the
cross filter K-correction $K_{Bi}(z)$, between apparent
magnitude $i$ and absolute magnitude $B$ \citep[see e.g.][]{Hogg02}
\be 
M^i_{\rm B}=i - DM(z) - K_{Bi}(z)
\label{eqn:MB}
\ee 
where $DM(z)$ is the distance modulus. We compute $K_{Bi}(z)$ from
the SDSS composite quasar spectrum of \citet{vanden01} and the
Johnson-B and SDSS $i$ filter curves.

The number density of quasars brighter than the flux limit $i^{\prime}$ is 
an integral over the luminosity function
\be
n(z,i<i^{\prime})=\int_{M^{\rm bright}_{\rm B}}^{M^{i^{\prime}}_{\rm B}}dM_{\rm B}
\Phi(M_{\rm B},z),
\ee
where $M^{\rm bright}_{\rm B}(z)$ and $M^{i^{\prime}}_{\rm B}(z)$ are the absolute
magnitudes corresponding to the bright and faint end apparent magnitude limits
as per eqn.~(\ref{eqn:MB}).

To check our model, we compute the cumulative number magnitude counts $n(<i)$
for the redshift range $0.3 < z < 3.0$ and compare to the measurement over the 
same redshift interval by \citet{vanden05}. We find that our 
model slightly overestimates the number counts of quasars. Specifically, 
\citet{vanden05} measured $n(<18.5) = 3.74~{\rm deg^{-2}}$ whereas our model 
predicts  $n(<18.5) = 4.63~{\rm deg^{-2}}$. We thus scale our model luminosity 
function down by this ratio so as to give the correct number counts for $i<18.5$.

\subsection{Estimating the Correlation Function}

If it is true that quasars depart from the gravitational clustering
hierarchy because of dissipative encounters, this would be expected to
occur for separations smaller than some length scale characteristic of
tidal effects or mergers. Proper rather than comoving coordinates
are appropriate for such an investigation. However, clustering
measurements are typically carried out in comoving coordinates, as
these are most intuitive in the linear regime where objects are sill
moving with the Hubble flow. We will compute the correlation
function in both proper and comoving units, but for definiteness
we use comoving units in the equations which follow. 

We consider quasars with a maximum velocity difference of 
$|\Delta v|<2000 \kms$, thus 
we will measure the redshift space correlation function projected
over this velocity interval 
\be 
w_{\rm p}(R,z)= \int_{-\frac{v_{\rm
      max}}{a H(z)}}^{\frac{v_{\rm max}}{a H(z)}} \xi_s(R,s,z)ds
\label{eqn:wp}
\ee
where $\xi_{s}$ is the quasar correlation function in redshift space, 
$v_{\rm max}=2000\kms$, $H(z)$ is the expansion rate at redshift $z$, and the
factor of $a=1/(1+z)$ converts the redshift distance to comoving units.

The redshift space correlation function, $\xi_s(R,s,z)$ is the
convolution of the velocity distribution in the redshift direction,
$F_v(v_z)$, with the real space correlation function $\xi(r,z)$, \be
\xi_s(R,s,z)=\int_{-\infty}^{\infty}\xi\left(\sqrt{R^2+x^2},z\right)
F_v(H(z)[x-s])dx.  \ee The radial velocity distribution $F_v(v_z)$ has
contributions from both peculiar velocities and
uncertainties in the systemic redshifts of the quasars. Provided that
the distance in redshift space over which we project 
contains most of the area under $F_v$, it is a good
approximation to replace the redshift space correlation function,
$\xi_s$, under the integral in eqn.~(\ref{eqn:wp}) with the real space
correlation function, $\xi$, since radial velocities will simply move
pairs of points within the volume.

The small number of close pairs will limit the number of bins in $R$
for which we can measure the correlation function. Since $w_{\rm p}$
may change significantly over these large bins, we choose to measure
the volume averaged projected correlation function integrated over each 
radial bin. We denote this dimensionless quantity by ${\bar W}_{\rm
  p}$.  Within a comoving bin $[R_{\rm min},R_{\rm max}]$ it can be
written as: 
\be 
{\bar W}_{\rm p}(R_{\rm min},R_{\rm max},z) \frac{\int_{-\frac{v_{\rm max}}{a H(z)}}^{\frac{v_{\rm max}}{a
      H(z)}}\int_{R_{\rm min}}^{R_{\rm max}} \xi(R,s,z)2\pi R dR
  ds}{V_{\rm shell}}
\label{eqn:wbar}\nonumber
\ee
where $V_{\rm shell}$ is the volume of the cylindrical shell in  redshift space
\be
V_{\rm shell}= \pi(R^2_{\rm max}-R^2_{\rm min})\left(\frac{2v_{\rm max}}{aH(z)}\right). 
\label{eqn:vbin}
\ee

The correlation function of a point process is computed by comparing
the number of pairs detected to the number expected in the absence
of clustering, taking into account the sample limits. We choose the 
estimator
\be
1+{\bar W}_{\rm p}(R_{\rm min},R_{\rm max})= 
\frac{\langle {\rm QQ}\rangle}{\langle{{\rm QR}\rangle}}.  \label{eqn:QQ}
\ee
Usually, random catalogs are constructed to determine the average
number of data-random pairs $\langle {\rm QR}\rangle$. However a
subtlety arises in the current context because for $\Delta \theta <
60\arcsec$, we have targeted the companion quasars (i.e the quasar
discovered from follow-up observations) to fainter magnitudes than the
quasars from the parent spectroscopic sample. Rather than estimate 
the correlation function from the number of pairs, we use the
number of \emph{companions}.  Specifically, $\langle {\rm
  QQ}\rangle$ is defined to be the number of companions
about quasars in the parent sample in a given comoving transverse
radial bin $[R_{\rm min},R_{\rm max}]$, and $\langle {\rm
    QR}\rangle$ is the average number of random companions in
this radial bin in the absence of clustering. In the case where the 
parent and companion samples are distinct (as is the case for the 
lens and $\chi^2$ samples) then  $\langle {\rm
    QQ}\rangle$ is just the number of binaries in the bin in question. However, 
if the parent and companion samples are identical (as is the case for the overlap
and spectroscopic samples), then  $\langle {\rm QQ}\rangle$ is \emph{twice}
the number of pairs, since each of the two quasars in the parent sample have a companion. 

Our model of the luminosity function in the previous section is used
to compute the average number of random companions ${\rm R}$ 
about the quasars ${\rm Q}$ in each parent quasar sample, taking into 
account the flux limits and various sources of incompleteness. We separately 
compute quasar-random contribution for each selection algorithm used to define our
clustering sample and we then add the results.  Specifically, for a comoving transverse 
radial bin $[R_{\rm min},R_{\rm max}]$, centered on $R$, we can write
\be 
\langle {\rm QR}\rangle= \sum_{j}^{N_{\rm qso}}n(z_j,i<i^{\prime})V_{\rm shell}S(z_j,\theta_j), 
\label{eqn:QR}
\ee
where $n(z,i<i^{\prime})$ is the number density of quasars above the
flux limit $i^{\prime}$, $V_{\rm shell}$ is the volume of each bin
in eqn.~(\ref{eqn:vbin}), and the sum is over all quasars in the parent
sample.  The quantity $S(z_j,\theta_j)$ is the
selection probability of detecting a companion about the $j$th
quasar. It is a function of angle and redshift since these quantities
parameterize our various selection algorithms, as discussed in
\S~\ref{sec:selection}.  Here $\theta_j=R/D(z)$ is the angle onto which the
(logarithmic) center of the bin projects for a quasar at
redshift $z$, and $D(z)$ is the comoving distance.  


In what follows we will compare our measured correlation function to
previous larger scale ($1~\hMpc-30~\hMpc$) clustering measurements
extrapolated as a power law down to the scales probed by our binaries.  Since ${\bar
  W}_{\rm p}$ in eqn.~(\ref{eqn:wbar}) is a function of redshift, we
must average it over the redshift distribution of our quasar sample
before a comparison can be made to the measurement from eqn.~(\ref{eqn:QQ}),
\be
   {\bar W}_{\rm p}(R_{\rm min},R_{\rm max}) = \frac{1}{N_{\rm qso}}
   \sum_j^{N_{\rm qso}}{\bar W}_{\rm p}(R_{\rm min},R_{\rm max},z_j)\label{eqn:wbarz}
\ee

PMN found that $\xi(r)$ is well fit by a power law, with  $\gamma=1.53$ and 
$r_0=4.8~\hMpc$, for the 2dF sample taken as a whole. They also 
found $r_0$ increases with redshift, from $r_0=3.4-5.9~\hMpc$, which 
we take into account in eqn.~(\ref{eqn:wbarz}).



In Figure~\ref{fig:Nc} we show the prediction for the
number of companions in the bin $0<R_{\rm prop}<50~\hkpc$ with $|\Delta
v|<2000~\kms$ as a function of redshift
\be
N_c=n(z,i<21)V_{\rm shell}[1 + {\bar W}_{\rm
    p}(0,50~\hkpc,z)].
\label{eqn:Nc}
\ee
For only this figure, we ignore the redshift evolution of the
correlation length and set $r_0=4.8~\hMpc$.  The expected mean number
of companions per quasar with $R<50~\hkpc$ is $\sim 2-3 \times
10^{-5}$ without incompleteness, so that in a sample of $\sim 50,000$
quasars we expect to 
find roughly $\sim 1$ quasar pair with separation $<50~\hkpc$. However, our
clustering sample already contains 20 quasars with transverse
separations this small.  Although our survey is far from complete,
there is already evidence for excess clustering over what is expected
from extrapolating the correlation function power law. In the next two sections
we will make this argument more precise.

\subsection{Computing the Selection Function}

To estimate the correlation function, we will compute the total number
of random pairs expected from eqn.~(\ref{eqn:QR}) by summing over the
redshifts of the quasars in each parent sample around which we searched
around for candidate companions. However, we must first consider
several sources of incompleteness. For the $\chi^2$ selected sample,
our $\chi^2<20$ cut results in a completeness fraction,
$C(z,\chi^2<20)$, shown in Figure~\ref{fig:chi2_complete}. In
addition, only a fraction of the pair candidates which satisfy the
criteria for our lens and $\chi^2$ selection algorithms have been
observed to date, and the fraction of candidates observed varies with
angular separation, because of our tendency to observe small
separation pairs first. Finally, for the $\theta > 60\arcsec$ pairs
which we find in the SDSS spectroscopic sample, the completeness
fraction is just that of the SDSS quasar survey and does not vary with
angle.

The selection probability in eqn.~(\ref{eqn:QR}) 
can therefore be written 
\be
S(z,\theta) = \left\{
\begin{array}{ll}
  F_{\rm lens}\left(\theta\right) & \mbox{$\theta \leq 3\arcsec$}\\
  & \\
  F_{\chi^2}(\theta)C\left(z |~\chi^2 < 20\right) & 
  \mbox{$3\arcsec < \theta \leq 60\arcsec$}\\
  & \\
  F_{\rm spec}  & \mbox{$\theta > 60\arcsec$}
\end{array} \right.\label{eqn:S}
\ee 
where $F_{\rm lens}(\theta)$, $F_{\chi^2}(\theta)$, and $F_{\rm
  spec}$ are the completeness fractions of the lens, $\chi^2$, and
spectroscopic samples, respectively. 

We take  $F_{\rm spec}=0.95$ following \citet{vanden05}.  The angular
selection functions, $F_{\rm lens}(\theta)$ and 
$F_{\chi^2}(\theta)$, of the lens and $\chi^2$ selection
algorithms, are computed by comparing, in bins, the number of pair candidates
which have been observed to date to the total number of candidates. 

For $\theta < 15\arcsec$, we choose the bin spacing such that 
each bin contains at least 6 objects. For $15\arcsec < \theta
< 60\arcsec$ we use ten logarithmically spaced angular
bins. 

The selection probability for the angular bin 
$[\theta_k,\theta_{k+1}]$ is 
\be
F_k=\frac{N_{\rm observed}}{N_{\rm observed}+N_{\rm remaining}}. \label{eqn:Fk}
\ee

There is an uncertainty in the number of remaining candidates for the
lens selection, which is based on the goodness of fit of a
multi-component PSF to the images of SDSS quasars \citep{Pindor03,Inada03}. 
Because our observations were heavily
biased towards those candidates which were likely to be quasar pairs, we
bracket the angular selection function $F_{\rm lens}(\theta)$ with
upper and lower limits corresponding to pessimistic and optimistic
assessments of the number of remaining candidates which are likely to
be confirmed as binaries. The lens candidates are given a grade of
A, B, or C determined by the goodness of fit of the multi-component
PSF. For the lower limit on $F_{\rm lens}(\theta)$, $N_{\rm
  remaining}$ is taken to be all candidates, whereas for the upper
limit $N_{\rm remaining}$ is restricted to only the candidates which
received a grade of A.  Note that to be conservative, we have 
ignored the intrinsic incompleteness of the lens selection algorithm, 
which is a function of separation, magnitude, and flux ratio \citep{Pindor03} of
the pair, and focus only on the incompleteness of our observations of the 
candidates.  

The color-color diagrams of the candidates identified by $\chi^2$
selection, were visually inspected prior to observations candidates
which overlapped the stellar locus were given a lower priority.
However, some interlopers were observed depending on a variety of
criteria.  For instance, we were more likely to observe pairs with
particularly small angular separations, or redshifts $z>2$ (because of
their use for Ly$\alpha$ forest studies), or those with particularly
bright magnitudes. Including the stellar locus interlopers identified
by the $\chi^2$ statistic as part of $N_{\rm remaining}$ in
eqn.~(\ref{eqn:Fk}) would overestimate the number of candidates likely
to be quasar pairs and hence underestimate our selection function. To
this end, we apply additional criteria to these remaining candidates
to filter out color-matches which are likely to be quasar-star
pairs. However, all quasar-star pairs we confirmed spectroscopically
are included as part of $N_{\rm observed}$ in eqn.~(\ref{eqn:Fk}).  In
this way, we conservatively err on the side of overestimating our
selection completeness, or underestimating the correlation function.

In addition to the criteria in eqn.~(\ref{eqn:cuts}), we only include
$\chi^2$ candidates which either (A) have colors consistent with the
UV-excess region of color space ($u-g< 0.6$) but
outside the region populated by white dwarfs \citep{qsoselect}, (B)
matched a member of the faint photometric catalog of
\citet{Richards04} (extended to the DR3 region), or (C) were optically
unresolved and matched a FIRST radio source.  Objects that satisfy
these criteria have a $\gtrsim 90\%$ probability of being a quasar.
For the UV-excess criteria, we conservatively required that the
candidates' 1$\sigma$ photometric errors left it inside the UV-excess
region and outside the white dwarf region. 



\begin{figure}
  \centerline{
    \epsfig{file=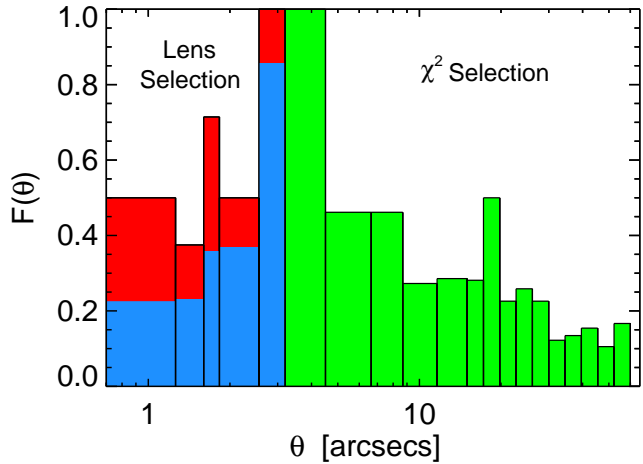,bb=40 0 504 360,width=0.5\textwidth}}
  \caption{ Angular selection function of binary quasars with $\theta
    < 60\arcsec$.  We have pasted $F_{\rm lens}(\theta)$ and
    $F_{\chi^2}(\theta)$ together at $\theta=3\arcsec$.  For $\theta
    \leq 3\arcsec$, the red and blue histograms represent, the upper
    and lower limits on the fraction of lens pair candidates
    observed to date, respectively. The green histogram shows the
    fraction of $\chi^2$ pair candidates observed so far.
    \label{fig:select}}
\end{figure}

The angular selection function of our binary quasar survey is shown in
Figure~\ref{fig:select}. 




\subsection{Excess Clustering}
\begin{figure*}
  \centerline{
    \epsfig{file=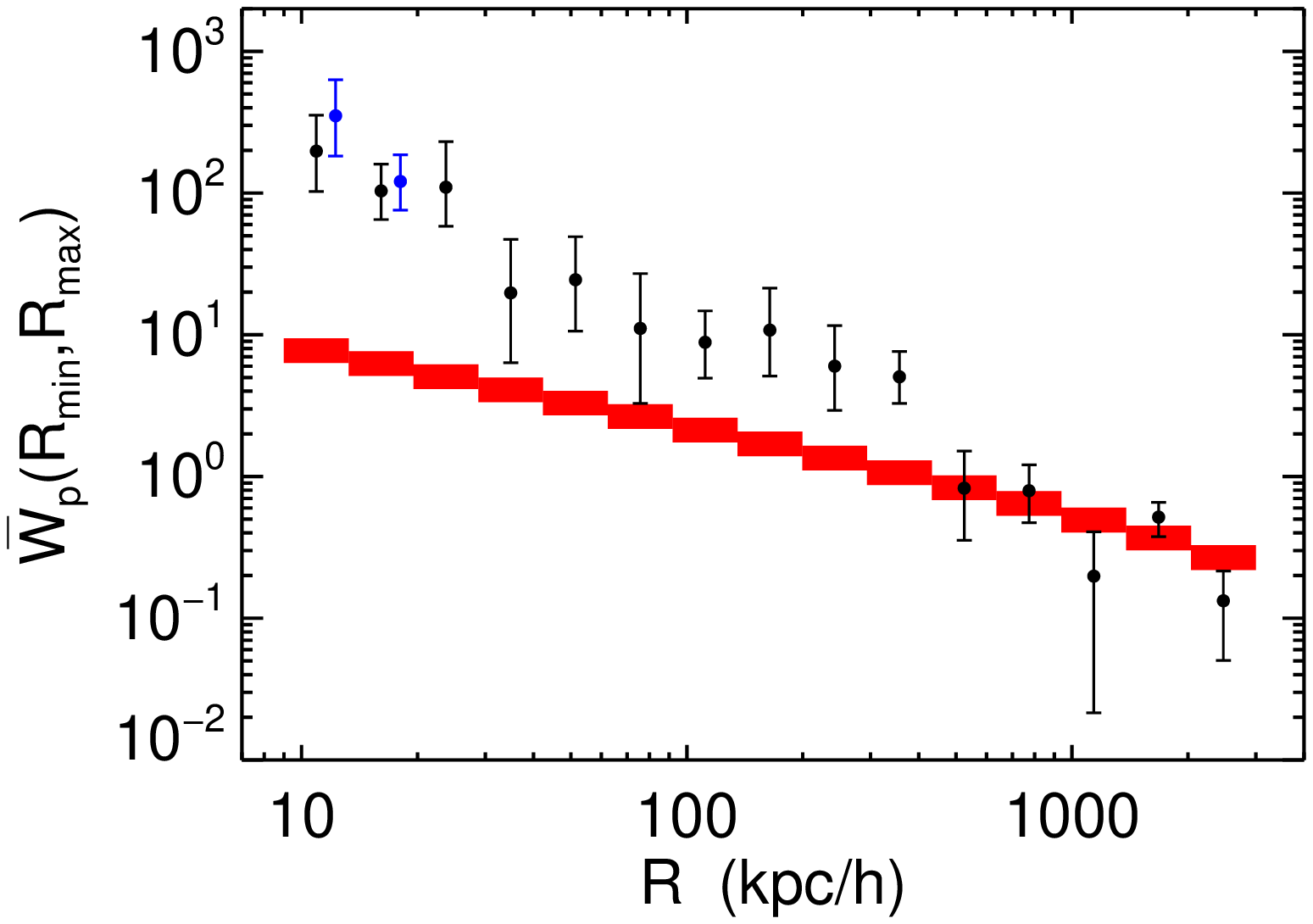,bb=82 72 576 412,width=0.5\textwidth}    
    \epsfig{file=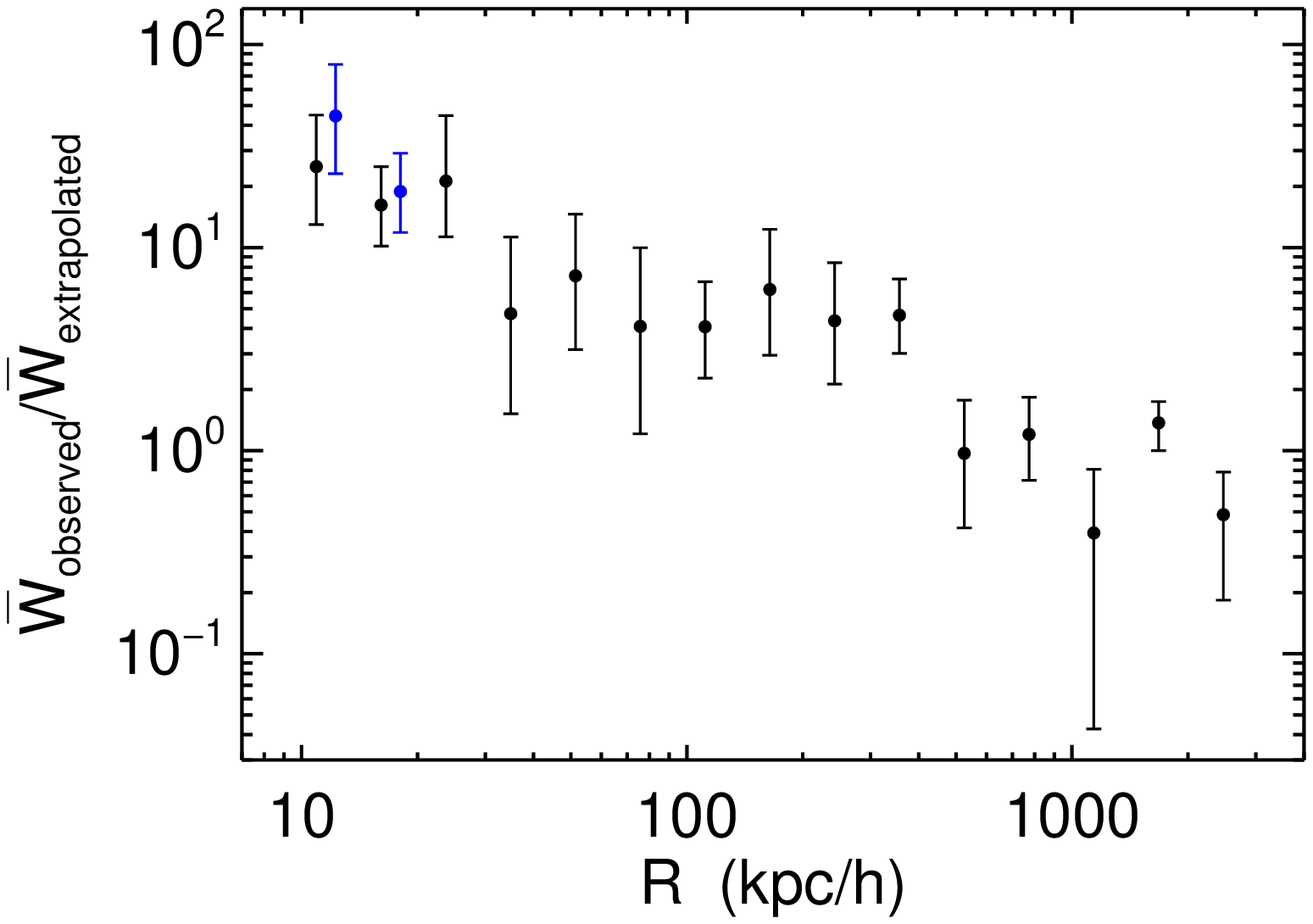,bb=82 72 576 412,width=0.5\textwidth}}
  \caption{ Small scale quasar clustering in proper coordinates
    \emph{Left:} Comparison of the projected correlation function
    ${\bar W}_{\rm p}(R_{\rm min},R_{\rm max},z)$ (see
    eqn.~\ref{eqn:wbar}) measured from our clustering sample with the
    prediction of the large scale measurement of PMN, extrapolated
    as a power law down to the scales probed by our binaries.  Error bars are one
    sigma Poisson counting errors. The blue points show the
    measurement of the projected correlation function if we use the
    lower limit (optimistic) for the selection function of our lens
    algorithm (blue histogram in Figure~\ref{fig:select}) to predict
    the number of quasar-random pairs, rather than the upper limit, 
    which are shown by the black points.  The blue
    points are offset slightly to the right for the sake of
    illustration. Red rectangles indicate the prediction 
    based on the large scale measurement of PMN, where the 
    height of the rectangles indicate the range of predictions
    based on one sigma errors in the correlation length measurements, 
    and the width indicate the bin used for each measurement. \emph{Right}
    Ratio of the projected correlation function to the (best-fit) prediction of 
    PMN. 
    \label{fig:clust_phys}}
\end{figure*}

\begin{figure*}
\centerline{
  \epsfig{file=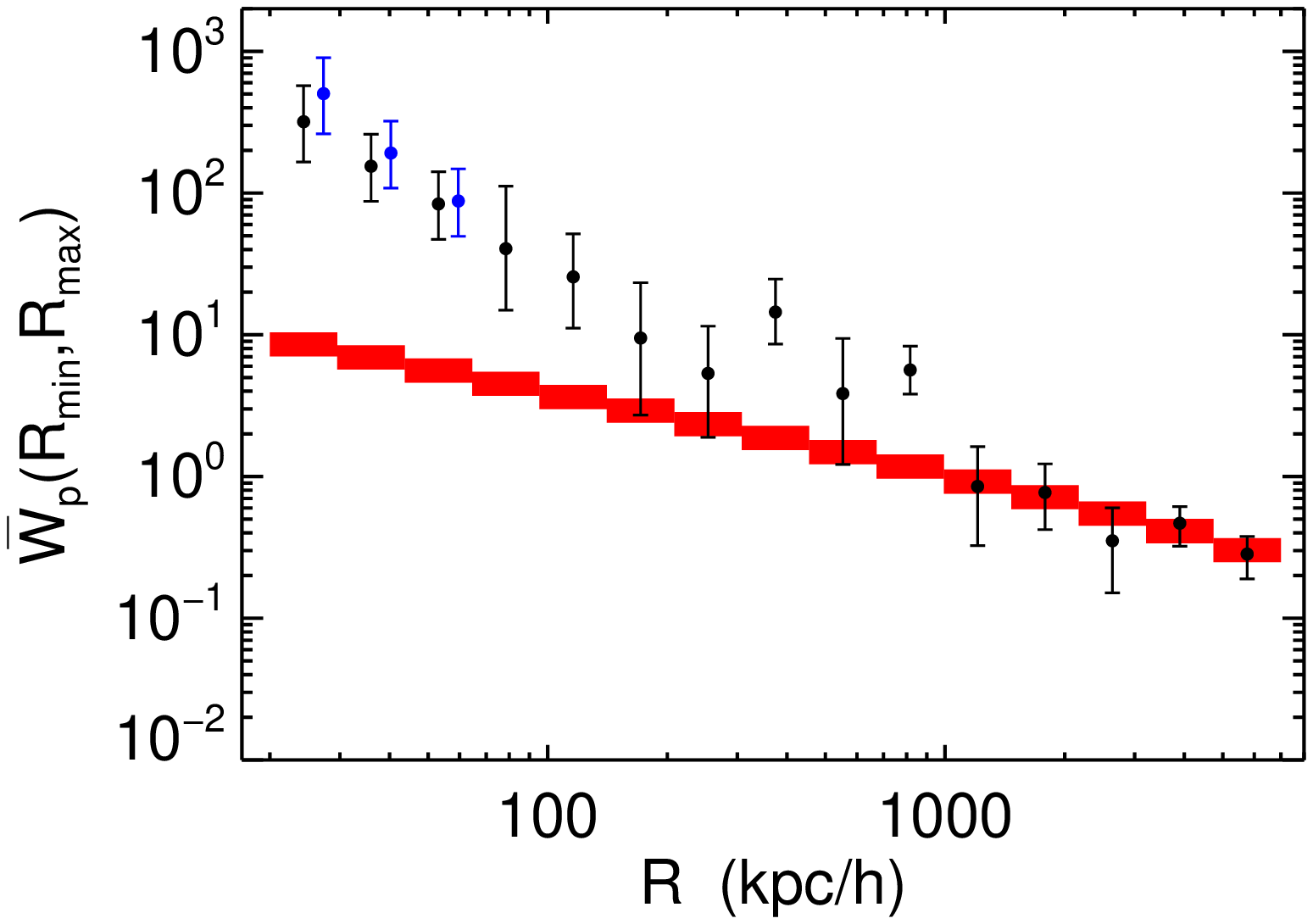,bb=82 72 576 412,width=0.5\textwidth}    
  \epsfig{file=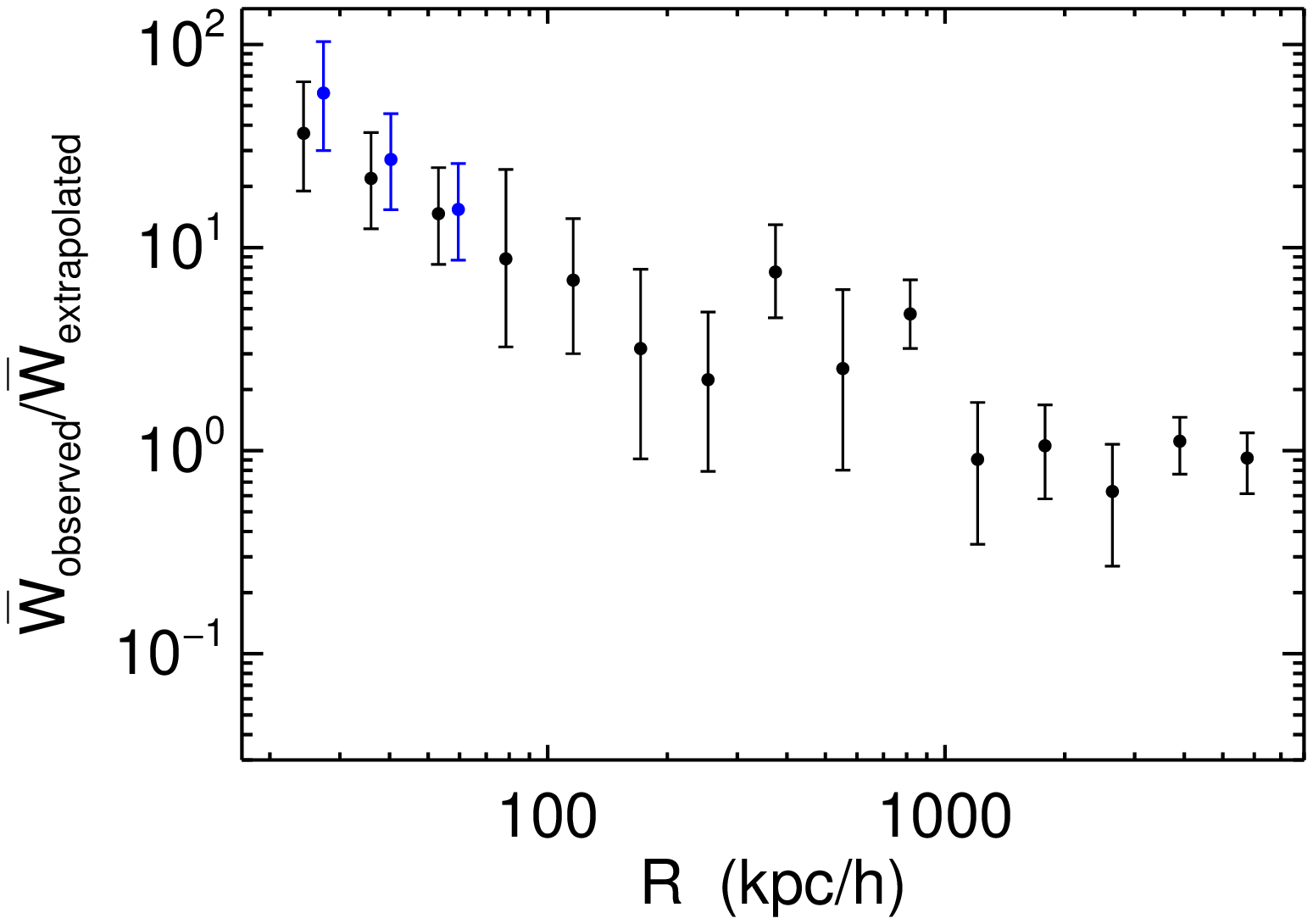,bb=82 72 576 412,width=0.5\textwidth}}
\caption{ Same as Figure~\ref{fig:clust_phys} except in comoving units. 
\label{fig:clust_comove}}
\end{figure*}

The individual quasar-random contributions for each selection
algorithm were computed by carrying out the
sum in eqn.~(\ref{eqn:QR}) over the respective parent
samples. Specifically, we summed over $39,142$  
quasars which were the parent sample for the lens sample, 
the $59,608$ (SDSS+2QZ quasars) which were the parent sample 
of the $\chi^2$ algorithm, and the 52,279 (SDSS only) quasars which made
up the parent sample of the spectroscopic sample. 

For the proper (comoving) projected correlation functions, we binned
our clustering sample into 15 logarithmically spaced bins in the range
$9~\hkpc < R_{\rm prop} < 3~\hMpc$ ($20~\hkpc < R < 7~\hMpc$). Both projected
correlation functions are compared to an extrapolation of the large
scale power law measured by PMN in the left panels of
Figures~\ref{fig:clust_phys} and \ref{fig:clust_comove}. The PMN
correlation function is in comoving coordinates, which was translated
to proper coordinates in each element of the average in
eqn.~(\ref{eqn:wbarz}).  The error bars are one sigma Poisson counting
errors, where we used the fitting formula in \citet{Gehrels86} for
$N<30$. The blue points show the measurement of the projected
correlation function if we use the lower limit (optimistic) for the
selection function of our lens algorithm (blue histogram in
Figure~\ref{fig:select}) to predict the number of quasar-random pairs,
while the black points use the upper limit. Uncertainty
in our selection function changes only the innermost few bins, because
our lens algorithm is restricted to angular separations $\theta <
3\arcsec$. These blue points are offset slightly to the right for the
sake of illustration. The red rectangles indicate the prediction for
the projected correlation function based on the large scale
measurement of PMN, extrapolated as a power law down to the scales probed by our
sample.  The heights of the rectangles are the range of predictions
based on the one sigma errors on the correlation lengths (for each
redshift interval) published by PMN, and the widths of the rectangles
indicate the bin used for each measurement.  The ratio of the measured
projected correlation functions to the extrapolation of the best fit
prediction of PMN are shown in the right panels of
Figures~\ref{fig:clust_phys} and \ref{fig:clust_comove}.

A summary of the data plotted in these Figures is given in
Tables~\ref{table:clust_phys} and \ref{table:clust_comove}. In the
Tables we compare our calculation for the number of random pairs in
each bin $\langle QR\rangle$ (eqn.~\ref{eqn:QR}) for our survey
to $\langle QR\rangle_{\rm perfect}$, which is the expected number of
random pairs for a `perfect' survey with no sources of incompleteness.
A comparison of these two quantities is a measure of our
incompleteness in each bin.

Our clustering measurement agrees with the measurements of PMN for
large proper (comoving) separations $R_{\rm prop}\gtrsim 400~\hkpc$
($R\gtrsim 1~\hMpc$), where the PMN measurement is valid. However, we
detect a significant excess over the expectation from extrapolating
the power law of the correlation function to smaller scales. Although
the uncertainties are large because of shot noise (and our uncertain
selection function for $\theta< 3\arcsec$), the excess is an order of
magnitude for proper (comoving) separations $R_{\rm prop}\lesssim 40~\hkpc$
($R\lesssim 100~\hkpc$). The excess is largest ($\sim 30$) on the
smallest scales $R_{\rm prop}\sim 15~\hkpc$ ($R\sim 30~\hkpc$). 

Our clustering statistic $W_{\rm p}(R_{\rm min},R_{\rm max})$ is the
average over a cylinder shell with velocity extent $4000~\kms$,
corresponding to a typical extent $r_{\rm prop}\sim 18~\hMpc$ ($r \sim
r_0)^{-\gamma}$, $W_{\rm p}(R_{\rm min},R_{\rm max})$ progressively 
flattens toward small scales, as is seen in red rectangles in
the left panels of Figures~\ref{fig:clust_phys} and
\ref{fig:clust_comove}. In contrast, our measured $W_{\rm p}(R_{\rm
  min},R_{\rm max})$ does not flatten, but rather resembles a power
law \emph{in projection}. This necessarily implies that the true three
dimensional correlation function of quasars is not a power law
$\xi(r)$, but rather gets \emph{progressively steeper on small
  scales.}

\begin{deluxetable*}{cccccccc}
\tablecolumns{8}
\tablewidth{0pc}
\tablecaption{Projected Correlation Function in Proper Coordinates\label{table:clust_phys}}
\tablehead{$R_{\rm min}$ & $R_{\rm max}$ & $\langle QQ\rangle$ & $\langle QR\rangle$ & $\langle QR\rangle_{\rm perfect}$ & ${\bar W}_{\rm p}$ & ${\bar W}^{\rm PMN}_{\rm p}$ & Ratio}
\startdata
\phn\phn\phn9.00  &\phn\phn 13.26  &  \phn\phn 4  &  0.02014 (0.01136)  &   0.03147  &   197.60 (350.98) &  7.89  &   25.0 (44.5)\\
\phn\phn13.26     &\phn\phn 19.53  &  \phn\phn 7  &  0.06694 (0.05749)  &   0.06922  &   103.57 (120.76) &  6.40  &   16.2 (18.9)\\
\phn\phn19.53     &\phn\phn 28.76  &  \phn\phn 5  &  0.04506            &   0.2240   &   109.96          &  5.17  &   21.3\\
\phn\phn28.76     &\phn\phn 42.37  &  \phn\phn 2  &  0.09635            &   0.4860   &   19.76           &  4.18  &    4.73\\
\phn\phn42.37     &\phn\phn 62.40  &  \phn\phn 3  &  0.11784            &   1.055    &   24.46           &  3.37  &    7.26\\
\phn\phn62.40     &\phn\phn 91.92  &  \phn\phn 2  &  0.1654             &   2.288    &   11.09           &  2.71  &    4.09\\
\phn\phn91.92     &\phn    135.39  &  \phn\phn 6  &  0.6090             &   4.964    &    8.85           &  2.17  &    4.07\\
\phn135.39        &\phn    199.42  &  \phn\phn 7  &  0.5938             &   10.77    &   10.79           &  1.74  &    6.22\\
\phn199.42        &\phn    293.74  &  \phn\phn 8  &  1.142              &   23.37    &    6.01           &  1.38  &    4.36\\
\phn293.74        &\phn    432.67  &  \phn    20  &  3.305              &   37.30    &    5.05           &  1.09  &    4.64\\
\phn 432.67       &\phn    637.31  &  \phn    23  &  12.59              &   13.25    &    0.83           &  0.85  &    0.97\\
\phn 637.31       &\phn    938.74  &  \phn    49  &  27.31              &   28.75    &    0.79           &  0.66  &    1.20\\
\phn 938.74       &       1382.73  &  \phn    71  &  59.26              &   62.38    &    0.20           &  0.50  &    0.39\\
1382.73           &       2036.71  &   195        &  128.6              &   135.3    &    0.52           &  0.38  &    1.37\\
2036.71           &       3000.00  &   316        &  279.0              &   293.6    &    0.13           &  0.27  &    0.48\\
\enddata
\tablecomments{ \footnotesize Data for clustering measurements shown in Figure~\ref{fig:clust_phys}. The 15 logarithmically spaced bins 
  of proper transverse separation are given by $R_{\rm min}$ and $R_{\rm max}$ in $\hkpc$. The number of observed pairs in our clustering sub-sample
  for each bin is given by $\langle QQ\rangle$. Our calculation of the number of random pairs in each bin is
  $\langle QR\rangle$ (eqn.~\ref{eqn:QR}), and $\langle QR\rangle_{\rm perfect}$ is the expected number of random pairs 
  for a `perfect' survey with no sources of incompleteness. Our measurement of the projected correlation function is given by 
  ${\bar W}_{\rm p}(R_{\rm min},R_{\rm max})$. The quantity ${\bar W}^{\rm PMN}_{\rm p}(R_{\rm min},R_{\rm max})$ is the prediction 
  from the larger scale clustering measurements of PMN, extrapolated as a power law down to the scale probed by our binaries, and averaged
  over the redshift distribution of our parent quasar sample (see eqn.~\ref{eqn:wbarz}). The ratio of our measurement to the
  prediction from PMN is given in the last column. The quantities in parentheses are the measurements if we use the lower limit
  for the selection function of the `lens' algorithm (blue histogram in Figure~\ref{fig:select}) to predict 
  $\langle QR\rangle$, which corresponds to the blue points in Figure~\ref{fig:clust_phys}.}
\end{deluxetable*}


\begin{deluxetable*}{cccccccc}
\tablecolumns{8}
\tablewidth{0pc}
\tablecaption{Projected Correlation Function in Comoving Coordinates\label{table:clust_comove}}
\tablehead{$R_{\rm min}$ & $R_{\rm max}$ & $\langle QQ\rangle$ & $\langle QR\rangle$ & $\langle QR\rangle_{\rm perfect}$ & ${\bar W}_{\rm p}$ & ${\bar W}^{\rm PMN}_{\rm p}$ & Ratio}
\startdata
\phn\phn20.00  &\phn\phn29.56   & \phn\phn 4  &   0.01251  (0.007935) &  0.02452   &   318.69 (503.12)  &   8.73  &   36.5 (57.6)\\
\phn\phn29.56  &\phn\phn43.68   & \phn\phn 5  &   0.03213  (0.02595)  &  0.06007   &   154.61 (191.66)  &   7.06  &   21.9 (27.1)\\
\phn\phn43.68  &\phn\phn64.54   & \phn\phn 5  &   0.05895  (0.05623)  &  0.1674    &    83.81 (87.92)   &   5.71  &   14.7 (15.4)\\
\phn\phn64.54  &\phn\phn95.38   & \phn\phn 3  &   0.07226             &  0.3886    &    40.52           &   4.61  &   8.80\\
\phn\phn95.38  &\phn\phn140.95  & \phn\phn 3  &   0.1127              &  0.8487    &    25.63           &   3.71  &   6.91\\
\phn140.95     &\phn208.28      & \phn\phn 2  &   0.1907              &  1.853     &     9.49           &   2.98  &   3.18\\
\phn208.28     &\phn307.80      & \phn\phn 3  &   0.4731              &  4.047     &     5.34           &   2.39  &   2.24\\
\phn307.80     &\phn454.85      & \phn    10  &   0.6467              &  8.839     &    14.46           &   1.91  &   7.58\\
\phn454.85     &\phn672.16      & \phn\phn 5  &   1.032               &  18.75     &     3.85           &   1.52  &   2.54\\
\phn672.16     &\phn993.29      & \phn    21  &   3.160               &  29.00     &     5.65           &   1.20  &   4.71\\
\phn993.29     &   1467.84      & \phn    20  &  10.80                &  13.29     &     0.85           &   0.94  &   0.91\\
1467.84        &   2169.12      &  \phn   42  &  23.71                &  24.96     &     0.77           &   0.73  &   1.06\\ 
2169.12        &   3205.45      &  \phn   70  &  51.78                &  54.51     &     0.35           &   0.56  &   0.63\\ 
3205.45        &   4736.89      &        166  &  113.1                &  119.0     &     0.47           &   0.42  &   1.11\\ 
4736.89        &   7000.00      &        317  &  246.9                &  259.9     &     0.28           &   0.31  &   0.92\\  
\enddata
\tablecomments{Same as Table~\ref{table:clust_phys} except in comoving coordinates (Figure~\ref{fig:clust_comove}).}
\end{deluxetable*}

\section{Discussion}
\label{sec:qso_gal}

On large proper comoving scales $\gtrsim 1~\hMpc$, quasars at $z \sim
1.5$, similar to high redshift galaxies, are strongly biased, and have
nearly the same correlation length that galaxies do in the local
universe. In the previous section we argued that the correlation
function of quasars becomes significantly \emph{steeper} on sub-Mpc
scales. Does small scale galaxy clustering or (quasar-galaxy
clustering) show a similar trend? As we compare to previous clustering
measurements in this section, all distances quoted are comoving.



At low redshift, \citet{Zehavi04} measured the projected correlation
function of the nearby ($z\lesssim 0.2$) SDSS main galaxy sample
($L\sim L^{\ast}$), down to scales as small as $\sim 100~\hkpc$.
\citet{Eis05} measured the small scale clustering of Luminous Red
Galaxies ($L\sim 5-10L^{\ast}$) at intermediate redshift ($0.2
\lesssim z \lesssim 0.3$), down to scales of $\sim 300~\hkpc$, by
cross correlating them with $L^{\ast}$ galaxies.  Both of these
studies find small deviations from a power law $\gamma\sim -1.9$, and
the correlation function does steepen toward smaller scales. However,
if these correlation function power laws were extrapolated from large ($\sim
1~\hMpc$) scales to small, as we have done above for quasars, the
clustering `excess' would not be more than $\sim 30\%$. In comparison,
we have found a much larger clustering excess, of a factor of $\sim 7$
on scales $R\sim 100~\hkpc$ and $\sim 2-3$ on scales $R \sim
300~\hkpc$ (see Table~\ref{table:clust_comove}).  One could argue that
the proper comparison is with the square of the clustering excess
measured by \citet{Eis05} (see discussion below), since it is a
cross-correlation with less luminous galaxies.  However
\citet{Zehavi05} measured the autocorrelation of LRGs at $\sim
400~\hkpc$, and also found an excess smaller than $\sim 30\%$.


Another complementary set of low redshift observations which probe
clustering on small scales ($10 \hkpc-1\hMpc$) are the SDSS
galaxy-galaxy lensing studies \citep{GS02,Sheldon04}, which measure
the galaxy-dark matter cross-correlation function. No excess small
scale clustering over a power law was detected in these studies either.

A comparison with small scale galaxy clustering at high redshift is
clearly more relevant to our purposes. \citet{Adel03} and \citet{Coil04}
have measured the correlation function of galaxies at $z\sim 3$ and
$z\sim 1$, respectively, down to scales as small as $\sim 100-200
\hkpc$, and no small scale excess clustering was found.  As the
redshift ranges of these measurements bracket the redshift range
probed by our sample of binary quasars, we expect that galaxies in the
redshift desert $z\sim 1.4-2.5$, coeval with the bulk of our binary
quasars (see Figure~\ref{fig:scatter}), also do not show any
enhancement in small scale clustering.

Thus, as first suggested by \citet{Djor91}, quasars at $z\sim 1.5$
depart from the power law clustering hierarchy followed by galaxies,
both in the local universe and at high redshift.  Galaxy interactions
are often implicated as a means of triggering, fueling, or forming
active galactic nuclei \citep{TT72,BH96,Bahcall97}, and it has been
claimed by several authors that excess quasar clustering is the
hallmark of these dissipative interaction events
\citep{Djor91,Koch99,MWF99}.  However, if this is the case, a similar
small scale enhancement is to be expected in the quasar-galaxy
correlation function, as follows.

If the quasar-quasar correlation function is $\xi_{\rm QQ}$, and
quasars trace galaxy overdensities with a linear relative bias $b_{\rm
  Q}$ defined by $\delta_{\rm Q}=b_{\rm Q}\delta_{\rm G}$, then the
ratio of the quasar-quasar correlation to the galaxy-galaxy
correlation is $\xi_{\rm QQ}\slash\xi_{\rm GG}=b_{\rm Q}^2$, and the
ratio of the quasar-galaxy cross correlation function to galaxy-galaxy
clustering is $\xi_{\rm QG}\slash\xi_{\rm GG}=b_{\rm Q}$. Since we
have argued that small scale quasar clustering is enhanced relative to
galaxy clustering at the same redshift, we expect to see the square
root of that enhancement in the quasar-galaxy cross correlation.

It has long been known that quasars are associated with enhancements
in the distribution of galaxies
\citep{BSG69,YG84,YG87,BC91,LS95,SBM95,HGC98,HG98,SBM00}, and the
foregoing argument has led us to ask if there is a small scale excess
of galaxy pairs around quasars.  However, nearly all studies of
quasar-galaxy correlations have been restricted to low redshifts
($z\lesssim 0.6$) and therefore low-luminosity, and the only
correlation function measurements that probed the redshift range
$1\lesssim z\lesssim 3$ relevant for comparison to our sample of
binaries measured a marginal cross-correlation of quasars with
galaxies \citep{Teplitz99}, or none at all \citep{BC93,
  CS99,Infante03}. We note that the disagreement in the literature on
whether radio-loud quasars are located in richer environments
\citep{YG84,YG87,EYG91} or not \citep{Fisher96,MD01,FIH01,Wold01}, is
not an issue in the current context, since only only one close binary
in our clustering sub-sample, LBQS1429$-$008, has a radio loud member.

Historically, there has been significant scatter in the low redshift
measurements of quasar-galaxy correlations \citep[see][Table 1 for a
  compilation of recent studies]{BBW01} caused by heterogeneous quasar
samples, methodology, and imaging depths. Furthermore, the measurement
of a factor of $\sim 4$ enhancement of the number of companion
galaxies around quasars compared to the mean number around galaxies,
measured by the Hubble Space Telescope Studies of $\sim 40$ nearby 
AGN by \citet{Fisher96} and \citet{MD01}, has been called into question by
\citet{FIH01} as being the result of sample biases. We thus focus on
the most recent determinations of quasar galaxy correlations, albeit
at low redshifts $z\lesssim 0.3$, by the 2dF and SDSS; these
surveys have
samples of $\sim 10,000$ AGN surrounded by $\sim 100,000$ galaxies at
their disposal.

\citet{Croom03} and \citet{Wake04} measured the ratio $\xi_{\rm
  QG}\slash\xi_{\rm GG}=b_{\rm Q}$, in the 2dF and SDSS surveys
respectively, finding it to be consistent with unity on small and
large scales.  The \citet{Croom03} measurement extends down to $\sim
900~\hkpc$; whereas the \citet{Wake04} measurement probes down to
scales $\sim 400 \hkpc$.  Both of these studies cross-correlated
spectroscopic samples of galaxies with spectroscopic samples of AGN;
hence, they are limited to low redshift $z\lesssim 0.3$, because the
spectroscopic galaxy samples are shallow, and large scales, because of
fiber collisions. At $z\sim 0.2$ the fiber collision limit corresponds
to $\sim 160~\hkpc$ in the SDSS and $\sim 80~\hkpc$ in the 2dF.  What
is needed is a clustering study using photometric galaxies around
spectroscopic quasars which could extend to redshift $z\sim 0.5$ and resolve
the scales $R\sim 10~\hkpc$ of most interest to us for the excess
quasar clustering \citep[see][in preparation]{Serber04}.


\section{Summary and Conclusions}
\label{sec:conc}

We have presented a sample of 218 new quasar pairs with transverse
separations $R_{\rm prop}< 1~\hMpc$ over the redshift range $0.5 \le z
\le 3.0$. Of these, 65 have angular separations $\theta <60\arcsec$
below the SDSS fiber collision scale.  Our 26 new pairs with proper
separations $R_{\rm prop}< 50~\hkpc$ ($\theta < 10\arcsec$) more than
doubles the number of binaries known with splittings this small.  Although
these binaries were discovered with a variety of selection algorithms,
we defined a statistical sub-sample selected with homogeneous
criteria, and computed its selection function taking into account
sources of incompleteness. We presented the first measurement of the
quasar correlation function on the small proper (comoving) scales
$10~\hkpc-400~\hkpc$ ($20~\hkpc-1~\hMpc$). We detect an order of
magnitude excess clustering for proper (comoving) separations $R_{\rm
  prop}\lesssim 40~\hkpc$ ($R\lesssim 100~\hkpc$), which grows to
$\sim 30$ on the smallest scale probed by our sample, $R_{\rm
  prop}\sim 15~\hkpc$ ($R\sim 30~\hkpc$).

We reviewed recent small scale measurements of galaxy clustering and quasar-galaxy
clustering and discussed the results in relation to the excess small
scale quasar clustering that we measured.  The quasar-galaxy
correlation function of redshift $z\sim 1.5$ quasars should show a
small scale clustering enhancement with amplitude roughly the square
root of the enhancement detected here. However, existing studies of the 
environments of quasars at $z\sim 1.5$ have been plagued by small
sample sizes and lack the statistics to address the clustering
strength on the $\sim 100~\hkpc$ scales of interest. 

Deep imaging of the binaries published here will provide valuable
information about their environments. The detection of significant
overdensities of galaxies coeval with the binary would support the
idea that enhanced quasar activity is triggered by galaxy
interactions, and it might suggest that quasars at high redshift trace
the biased peaks which are the progenitors of the rich clusters we see
today \citep{ER88,CK89,NS93,Djor99a,Djor99b,Djor03}. Intriguingly,
\citet{Fuku04} took deep Subaru images of the $z=4.25$ quasar
discovered by \citet{Schneider00}, and found no evidence for an
overdensity of galaxies. Similar deep imaging studies of high redshift
quasar environments (for single quasars) have been conducted by
\citet{Infante03} at $z\sim 3$, and by Djorgovski and collaborators at
$z\gtrsim 4$ \citep{Djor99a,Djor99b,Djor03,Stia05}. The binaries in
our sample offer an opportunity to conduct analogous studies over a
range of lower and more accessible redshifts, with the added bonus
that one expects these extremely rare binary systems to trace even
richer environments.

Measurements of the shape of the quasar-quasar and quasar-galaxy
correlation function on the small scales probed by our binaries
will yield valuable insights into the physical processes that trigger
quasar activity and will help explain how quasars are embedded in the
structure formation hierarchy.  Reproducing the excess clustering with
semi-analytical models \citep{KH02} and halo models (see e.g. PMN) 
would provide constraints on the distribution of quasars in
dark matter halos. Another interesting question is whether the quasars
in these binaries have significantly longer lifetimes than `field'
quasars \citep{HH01,MW01}, which would have interesting
implications for the masses of supermassive black holes \citep{WL04}
in the richest regions of the Universe.

We close with the reminder that our survey for quasar pairs is ongoing
and less than 50\% complete. We expect to find a comparable number of
binaries in the current SDSS quasar sample.  Furthermore, the
faint photometric quasar selection techniques of \citet{Richards04} aim to
construct a sample of $\sim 10^6$ quasars.  Extrapolating from the
number of pairs published here, we would expect to find $\sim
1000$ new binaries in a sample of this size, which would allow a 
much more precise measurement of the correlation function on small scales. 


\acknowledgments 
JFH wishes to thank his thesis advisor David Spergel for advice,
guidance, and helpful discussions over the course of this work during
his time in Princeton.  For part of this study JFH was supported by
the Proctor Graduate fellowship at Princeton University and by a
generous gift from the Paul \& Daisy Soros Fellowship for New
Americans. The program is not responsible for the views expressed.
JFH is currently supported by NASA through Hubble Fellowship grant \#
01172.01-A awarded by the Space Telescope Science Institute, which is
operated by the Association of Universities for Research in Astronomy,
Inc., for NASA, under contract NAS 5-26555. Thanks to Bob
Nichol and Alex Gray for help with the faint photometric quasar
catalog.  We are grateful to the observing specialists at Apache Point
Observatory, Russet MacMillan, John Barentine, Bill Ketzeback, and
Jack Dembicky for many nights of hard work.

Funding for the creation and distribution of the SDSS
Archive has been provided by the Alfred P. Sloan Foundation, the
Participating Institutions, the National Aeronautics and Space
Administration, the National Science Foundation, the U.S. Department
of Energy, the Japanese Monbukagakusho, and the Max Planck
Society. The SDSS Web site is http://www.sdss.org/.  The SDSS is
managed by the Astrophysical Research Consortium (ARC) for the
Participating Institutions. The Participating Institutions are The
University of Chicago, Fermilab, the Institute for Advanced Study, the
Japan Participation Group, The Johns Hopkins University, the Korean
Scientist Group, Los Alamos National Laboratory, the
Max-Planck-Institute for Astronomy (MPIA), the Max-Planck-Institute
for Astrophysics (MPA), New Mexico State University, University of
Pittsburgh, University of Portsmouth, Princeton University, the United
States Naval Observatory, and the University of Washington.  

This paper is based in part on data collected at the W. M. Keck
Observatory, the Apache Point Observatory 3.5m Telescope, the
Hobby-Eberly Telescope, and the European Southern Observatory, Chile,
(proposal \#67.A-0544).  The Apache Point Observatory (APO) 3.5-meter
telescope, is owned and operated by the Astrophysical Research
Consortium.  The W. M. Keck Observatory is operated as a scientific
partnership among the California Institute of Technology, the
University of California, and the National Aeronautics and Space
Administration.  The Hobby-Eberly Telescope (HET) is a joint project
of the University of Texas at Austin, the Pennsylvania State
University, Stanford University, Ludwig-Maximillians-Universit\"at
M\"unchen, and Georg-August-Universit\"at G\"ottingen.  The HET is
named in honor of its principal benefactors, William P. Hobby and
Robert E. Eberly.  The Marcario Low-Resolution Spectrograph is named
for Mike Marcario of High Lonesome Optics, who fabricated several
optics for the instrument but died before its completion; it is a
joint project of the Hobby-Eberly Telescope partnership and the
Instituto de Astronom\'{\i}a de la Universidad Nacional Aut\'onoma de
M\'exico. 

DPS acknowledges support from NSF grant (AST03-07582) and MAS acknowledges
the support of NSF grant (AST-0307409). 


\begin{appendix}
  
  In this appendix we present the results of all of our follow up
  observations not included above, as well as list of projected pairs
  of quasars in the SDSS quasar sample which have similar
  redshifts. These tables will facilitate future studies of close
  quasar pairs and prevent the duplicate observations of candidates
  already observed by the SDSS or our follow up observation program.
  
  In Table~\ref{table:overlapsample} we list all projected pairs of
  quasars in the combined SDSS+2QZ spectroscopic catalog with proper
  transverse separations $R_{\rm prop}<1~\hMpc$ at the foreground
  quasar and redshift difference $\Delta z < 0.5$. All projected pairs
  of quasars discovered from our follow up spectroscopic observations are listed in
  Table~\ref{table:aposample}. Finally, Table~\ref{table:starsample}
  lists all of the quasar-star pairs identified from our follow up
  observations.  The full versions of these tables are published in 
  in the electronic version of this article.  
  
  \begin{deluxetable*}{lcccccccccc}
\tablecolumns{11}
\tablewidth{0pc}
\tablecaption{Projected Pairs of  Quasars Discovered in Overlapping Plates\label{table:overlapsample}}
\tablehead{Name & RA & Dec &$u$ &$g$ & $r$ & $i$ & $z$ & z & $\Delta \theta$ & $R_{\rm prop}$}
\startdata
\hfill SDSSJ0000$+$0055A &  	00:00:42.91 &  	$+$00:55:39.5 &  	18.30  &  	18.16  &  	17.99  &  	18.01  &  	17.88  &   0.95  &  
 170.7      &  	 961.9  \\
\hfill SDSSJ0000$+$0055B &  	00:00:42.91 &  	$+$00:55:39.5 &  	20.01  &  	19.69  &  	19.44  &  	19.39  &  	19.43  &   1.18  & 
            &	        \\
\hfill SDSSJ0004$+$0000A &  	00:04:42.18 &  	$+$00:00:23.4 &  	19.24  &  	19.09  &  	18.85  &  	18.95  &  	19.07  &   1.01  &  
 134.0  &  	 626.4  \\	
\hfill SDSSJ0004$-$0001B &  	00:04:42.18 &  	$+$00:00:23.4 &  	20.91  &  	20.39  &  	20.22  &  	19.83  &  	19.52  &   0.58  & 
            &	        \\
\enddata

\tablecomments{\footnotesize Quasars labeled SDSS or 2QZ are members
  of the SDSS or 2QZ spectroscopic quasar catalog.  The redshift of
  each quasar is indicated by the column z and the foreground quasar
  is always designated `A'. The column labeled $R_{\rm prop}$ is the
  transverse proper separation at the foreground quasar in $\hkpc$, extinction
  corrected SDSS five band PSF photometry is given in the columns $u$,
  $g$, $r$, $i$, and $z$, and $\Delta \theta$ is the angular separation in 
  arcseconds} 
\end{deluxetable*}

  \begin{deluxetable*}{lccccccccccc}
\tablecolumns{12}
\tablewidth{0pc}
\tablecaption{Projected Pairs of Quasars Discovered From Follow Up Spectroscopy\label{table:aposample}}
\tablehead{Name & RA & Dec &$u$ &$g$ & $r$ & $i$ & $z$ & z & $\Delta \theta$ & $R$ & $\chi^2$}
\startdata
\hfill APOJ0002$-$0053A &  	00:02:12.53 &  	$-$00:53:11.7 &  	20.72  &  	20.59  &  	20.34  &  	20.12  &  	20.17  &   1.54  &  
\phn7.3  &  	\phn44.5  &  	\phn20.9 	\\	
\hfill SDSSJ0002$-$0053B &  	00:02:12.53 &  	$-$00:53:11.7 &  	20.10  &  	19.64  &  	19.52  &  	19.41  &  	19.23  &   2.21  & 
&	            &		\\	
\hfill SDSSJ0036$-$1109A &  	00:36:49.63 &  	$-$11:09:29.8 &  	19.08  &  	18.85  &  	18.67  &  	18.45  &  	18.67  &   1.51  &  
\phn4.7  &  	\phn28.5  &  	\phn19.0 	\\	
\hfill APOJ0036$-$1109B &  	00:36:49.63 &  	$-$11:09:29.8 &  	20.92  &  	20.64  &  	20.46  &  	20.52  &  	20.24  &   2.18  & 
&	            &		\\	
\enddata

\tablecomments{Quasars labeled SDSS or 2QZ are members of the SDSS or
  2QZ spectroscopic quasar catalog.  Quasars discovered from follow up
  spectroscopy are labeled APO. The redshift of each quasar is
  indicated by the column z and the foreground quasar is always
  designated `A'. The column labeled $R_{\rm prop}$ is the transverse
  proper separation at the foreground quasar in $\hkpc$, extinction
  corrected SDSS five band PSF photometry is given in the columns
  $u$, $g$, $r$, $i$, and $z$, $\Delta \theta$ is the angular
  separation in arcseconds, and $\chi^2$ is the value of our color
  similarity statistic.}
\end{deluxetable*}

  \begin{deluxetable*}{lcccccccccc}
\tablecolumns{11}
\tablewidth{0pc}
\tablecaption{Quasar-Star Pairs Discovered From Follow Up Spectroscopy$^{\dagger}$
\label{table:starsample}}
\tablehead{Name & RA & Dec &$u$ &$g$ & $r$ & $i$ & $z$ & z & $\Delta \theta$ & $\chi^2$}
\startdata
\hfill SDSSJ0006$+$0026A &  	00:06:14.00 &  	$+$00:26:05.0 &  	20.85  &  	20.09  &  	19.98  &  	19.95  &  	19.81  &  
2.51  &  	 28.0  &  	\phn42.4 	\\	
\hfill APOJ0006$+$0026B &  	00:06:14.00 &  	$+$00:26:05.0 &  	21.04  &  	20.16  &  	19.83  &  	19.70  &  	19.78  &  
            &	            &		\\	
\hfill SDSSJ0015$+$0048A &  	00:15:57.08 &  	$+$00:48:22.4 &  	20.74  &  	19.95  &  	19.63  &  	19.57  &  	19.29  &  
2.31  &  	 53.5  &  	\phn\phn5.4 	\\	
\hfill APOJ0015$+$0048B &  	00:15:57.08 &  	$+$00:48:22.4 &  	21.12  &  	20.38  &  	20.08  &  	19.96  &  	19.87  &  
            &	            &		\\	
\enddata
\tablecomments{Quasars labeled SDSS or 2QZ are members of the SDSS or
  2QZ spectroscopic quasar catalog and are always designated by `A'.
  Stars discovered from follow up spectroscopy are labeled APO and
  designated by `B'.  The redshift of the quasar is indicated by the
  column z, extinction corrected SDSS five band PSF photometry is
  given in the columns $u$, $g$, $r$, $i$, and $z$, $\Delta \theta$ is
  the angular separation in arcseconds, and $\chi^2$ is the value of
  our color similarity statistic.}
\end{deluxetable*}

\end{appendix}

\end{document}